\documentclass[fleqn,usenatbib,useAMS]{mnras}

\usepackage{graphicx}	
\usepackage{amsmath}	
\usepackage{amssymb}	
\usepackage{multicol}        
\usepackage{bm}		
\usepackage{pdflscape}	

\newcommand{\kms}{\,km\,s$^{-1}$} 

\usepackage[T1]{fontenc}
\usepackage{ae,aecompl}

\usepackage{newtxtext,newtxmath}

\newcommand{\be}{\begin{equation}}
\newcommand{\ee}{\end{equation}}
\newcommand{\bea}{\begin{eqnarray}}
\newcommand{\eea}{\end{eqnarray}}

\title[Reliable Photometric Membership. I]{Reliable Photometric Membership (RPM) of Galaxies in Clusters. I. A Machine Learning Method and its Performance in the Local Universe}

\author[P. Lopes \& A. Ribeiro]{
Paulo A. A. Lopes$^{1}$\thanks{E-mail: plopes@astro.ufrj.br}, Andr\'e L. B. Ribeiro$^{2}$,
\\
$^{1}$Observat\'orio do Valongo, Universidade Federal do Rio de Janeiro, Ladeira do Pedro Ant\^onio 43, Rio de Janeiro, RJ, 20080-090, Brazil\\
$^{2}$Laborat\'orio de Astrof\'isica Te\'orica e Observacional -- Departamento de Ci\^encias Exatas e Tecnol\'ogicas --Universidade Estadual de Santa Cruz,\\ 45650-000, Ilh\'eus, BA, Brazil\\
}

\date{Accepted XXX. Received YYY; in original form ZZZ}

\pubyear{2019}

\begin{document}
\label{firstpage}
\pagerange{\pageref{firstpage}--\pageref{lastpage}}
\maketitle

\begin{abstract}
We introduce a new method to determine galaxy cluster membership based solely on photometric properties. We adopt a machine learning approach to recover a cluster membership probability from galaxy photometric parameters and finally derive a membership classification. After testing several machine learning techniques (such as Stochastic Gradient Boosting, Model Averaged Neural Network and k-Nearest Neighbors), we found the Support Vector Machine (SVM) algorithm to perform better when applied to our data. Our training and validation data are from the Sloan Digital Sky Survey (SDSS) main sample. Hence, to be complete to $M_r^* + 3$ we limit our work to 30 clusters with $z_{\text{phot-cl}} \le 0.045$. Masses ($M_{200}$) are larger than $\sim 0.6\times10^{14} M_{\odot}$ (most above $3\times10^{14} M_{\odot}$). Our results are derived taking in account all galaxies in the line of sight of each cluster, with no photometric redshift cuts or background corrections. Our method is non-parametric, making no assumptions on the number density or luminosity profiles of galaxies in clusters. Our approach delivers extremely accurate results (completeness, C $\sim 92\%$ and purity, P $\sim 87\%$) within R$_{200}$, so that we named our code {\bf RPM}. We discuss possible dependencies on magnitude, colour and cluster mass. Finally, we present some applications of our method, stressing its impact to galaxy evolution and cosmological studies based on future large scale surveys, such as eROSITA, EUCLID and LSST.
\end{abstract}




\section{Introduction}

Galaxy clusters represent the most massive and latest systems to form in the Universe due to their own gravity. Therefore, they provide an important cosmological tool, especially through the investigation of the cluster mass function and its time evolution, which display a strong dependence on several cosmological parameters \citep{eke98,bah03,deH16,man14,sch17,pla16}.

Due to their large gravitational potential clusters also provide important sites to study galaxy evolution. Galaxies in the local Universe are characterized by a morphology-density (and colour-density) relation, highlighting the importance of the environment for influencing their physical properties. Most galaxies in high density regions of the Universe (groups and clusters) are bulge dominated and show little star formation, being called early-type galaxies. Low density regions are dominated by blue spirals, with significant star formation rate, which are termed late type.
Those relations also have a strong dependence on mass, as red early-type objects are the most massive.

Many studies based on galaxy clusters (e.g., the morphology and colour-density relations or richness estimates) are obviously better performed if it is possible to reliably determine which galaxies are members of a cluster. Ideally that would be fairly well achieved with spectroscopic surveys, but those are rather observationally expensive. Although there are currently large spectroscopic surveys available, their volume is many times smaller than what is reached by photometric surveys. Hence, many groups and clusters have incomplete and shallow spectroscopic sampling or totally lack redshifts. Current deep all sky cluster catalogs (such as the one provided by the Planck satellite) or large deep photometric catalogs ({\it e.g.} KIDs or DES) already provide many more cluster candidates than is possible to probe spectroscopically. This situation will only get worse with future large area surveys, such as EUCLID, LSST and eROSITA.

When a spectroscopic follow-up is poor or absent, galaxy cluster population studies have been traditionally done with statistical corrections. Those can be derived from the subtraction of the background contribution (estimated from nearby regions) from the cluster counts. Although they provide an important tool they may be affected by systematic effects (such as cosmological variance), potentially biasing the results. If the background is over (or under) estimated, studies relying on fractions of galaxy populations become biased. Another drawback of the statistical correction approach is the fact they are only valid for combined samples, preventing any investigation on individual galaxies.

Hence, knowledge of galaxy membership in clusters is paramount for cosmology and galaxy evolution studies. However, only a few works have previously addressed the issue of robust membership assignment from photometric information \citep{geo11,bru00,roz15,cas16,bel18}. The latter actually provides a cluster finder that has the membership assignment as part of the process.

In the current work we develop a new method based on a machine learning approach, relying on a Support Vector Machine (SVM) algorithm. We show that the use of simple galaxy photometric parameters (colours, magnitude, $z_{\text{phot}}$, concentration and radius), as well as environmental properties (radial offset and local density), result in extreme accurate estimates of cluster membership. We also verify that some of those parameters (such as photometric redshift and local density) are key for the classification procedure. However, their use alone (without taking account more parameters) leads to a poorer classification. We expect that more accurate photometric redshifts in the near future (from surveys such as the Javalambre Physics of the Accelerating Universe Astrophysical Survey, J-PAS) would give a greater contribution for this classification problem.

This paper is organized as follows: in $\S$2 we describe the cluster and galaxy samples. We also describe the galaxy properties that are used for training and validation. In $\S$3 we introduce the machine learning algorithms we tested in our data, as well as the feature choice, tune and model selection. We also describe the membership probabilities and define completeness and purity. Our main results are presented in $\S$4, while we show some applications in $\S$5. In $\S$6 we draw our conclusions. The cosmology assumed in this work is $\Omega_{\rm m}= $ 0.3, $\Omega_{\lambda}= $ 0.7, and H$_0 = 100$ $\rm h$ $\rm km$ $s^{-1}$ Mpc$^{-1}$, with $\rm h$ set to 0.7.

\section{Data}
\label{data} 

This work is based on two cluster samples. The first one is an optically selected sample (NoSOCS) described in \citet{lop09a,lop09b,lop14}. The second is composed of clusters selected in X-rays and through the Sunyaev-Zel'dovich effect (called ESZ+X-ray), being described in \citet{lop18}. 

Both samples are limited to $z \sim 0.1$. However, for
the current work we limit ourselves to clusters below $z_{\text{phot-cl}} = 0.045$ (the minimum and median redshifts are $z_{\text{phot-cl}} = 0.031$ and $0.037$, respectively). We do so as we want to sample systems that are complete to at least $M^*+3$ in the $r$-band. 

We give the main characteristics of the original samples below, but refer the reader to the above references for more details on these data sets. 

\subsection{The Cluster Samples} 
\label{cls_samples} 


The Northern Sky Optical Cluster Survey (NoSOCS) is originated from the digitized version of the Second Palomar Observatory Sky Survey (POSS-II; DPOSS, \citealt{gal04,ode04}). The supplemental version of NoSOCS \citep{lop04} goes deeper ($z \sim 0.5$), but covers a smaller region than the main NoSOCS catalog \citep{gal03, gal09}.

In \citet{lop09a,lop09b} we extracted SDSS photometric and spectroscopic data for all NoSOCS clusters of the supplemental version \citep{lop04} that were inside the SDSS DR5 footprint. We kept a subset of 127 systems, at $z \le 0.100$, with enough spectra for cluster redshift determination, as well as to perform a virial analysis. This low-redshift sample was complemented with 56 more massive systems from \citet{rin06}. The number of redshifts for those 183 clusters was later updated with DR7 data \citet{lop14}, when we also updated the code used to select members and estimate velocity dispersion, physical radius and mass.  Due to this change the sample was reduced to 155 clusters for which we have 6,415 member galaxies. Our clusters span 
the range $150 \la \sigma_P \la 950$ km s$^{-1}$, or the equivalently in terms of mass, $10^{13} \la M_{200} \la 10^{15} M_{\odot}$.


The second data set we consider in this work comes from two independent selected samples from \citet{and17}. One is the Planck Early Sunyaev-Zel'dovich (ESZ) sample of 189 SZ clusters, containing 164 clusters at $z < 0.35$. 
The other sample has 100 clusters from a flux-limited X-ray sample for which Chandra data are also available. From now on we call this combined data set as ESZ+X-ray. 

The combined ESZ and X-ray sample at low ($z \lesssim 0.1$) redshift has 72 clusters, with 30 objects in common to the two data sets. ESZ clusters are generally more massive than the X-ray objects, which in turn are also more massive than the NoSOCS. That can be seen in Fig. 9 of \citet{lop18}. The masses in the ESZ+X-ray sample range from $M_{200} \sim 0.2\times10^{14} M_{\odot}$ to $\sim 30\times10^{14} M_{\odot}$, with most objects having masses above $3\times10^{14} M_{\odot}$.

For the current work we consider only clusters with $z_{\text{phot-cl}} \le 0.045$, so that we end up having 18 NoSOCS clusters and 12 ESZ+X-ray systems. There are 9991 galaxies, with $M_r \le M_r^*+3$, within 4 Mpc from the center of NoSOCS clusters and 7785 for the ESZ+X-ray objects. The combined galaxy data set (combining NoSOCS and ESZ+X-ray) has 15802 galaxies, after eliminating common objects. These 30 clusters have masses in the range between $M_{200} \sim 0.6\times10^{14} M_{\odot}$ to $\sim 30\times10^{14} M_{\odot}$. However, the sample comprises fairly massive clusters (the mean mass is $M_{200} = 5.05\times10^{14} M_{\odot}$). Their photometric estimates of $R_{200}$ (see $\S$\ref{clsphot}) are between 0.66 Mpc and 1.91 Mpc.

\subsection{Galaxy photometry} 
\label{gal_phot}

The photometric data used in this paper comes from the SDSS DR7, except for a few galaxies in the ESZ+X-ray clusters without photometry in that data release (for which we use DR14). We did so for two reasons. First, galaxies for the NoSOCS clusters had already been selected from DR7 \citep{lop09a, lop14}. Second, as explained in \citet{lop18}, when selecting galaxies for the ESZ+X-ray systems we wanted the largest number of redshifts as possible, but decided to use the DR7 photometry (whenever available) as some bright galaxies have their fluxes underestimated in DR14. That is due to changes in the photometric pipeline after the DR8 \citep{dr8sdss2011}.

We have selected only objects from the "Galaxy view" (so that only \texttt {PRIMARY} objects are allowed) to avoid duplicate observations. We applied "standard flags" for clean photometry. All the magnitudes retrieved from SDSS are de-reddened model magnitudes. A {\it joined} query of the "Galaxy" and "SpecObj views" is performed, as we select imaging and spectra. That is also done to select photometric redshifts ($z_{\text{phot}}$) from the "Photoz table". The query returns all galaxies within 5 Mpc of each cluster centroid.

As described in the "Photometric Redshifts entry in Algorithms" in the SDSS {\it sky server}\footnote{http://skyserver.sdss.org/dr7/en/help/docs/algorithm.asp?key=photoz}, the $z_{\text{phot}}$ estimate \footnote{see also https://www.sdss.org/dr14/algorithms/photo-z/} is based on a machine learning approach, employing a kd-tree nearest neighbor fit (KF). To derive k$-$correction, galaxy types, distance modulus and other parameters for each galaxy, they combine the above method with a template fitting procedure. The estimated redshift error is of order $\sim 0.02$ \citep{bec16}).

\subsection{Galaxy spectroscopic redshifts} 
\label{gal_spec}

As mentioned above, we selected spectroscopic redshifts for the galaxies in our clusters in the same query ({\it joined} query of the "Galaxy" and "SpecObj views"). For the NoSOCS clusters, photometric and spectroscopic information for all galaxies are from the SDSS DR7, while for the ESZ+X-ray sample, spectra is from DR14 and photometry from DR7 (whenever possible). 

As the completeness of SDSS is affected by issues like fiber collision, we have also gathered additional redshifts from the NASA/IPAC Extragalactic Database (NED). As described in \citet{lop18}, we selected all galaxies in NED with a reliable spectroscopic redshift within 5 Mpc from the ESZ+X-ray sample. According to the quality code for the redshifts, we used the galaxies with the code listed as ``blank" (``usually a reliable spectroscopic value'') or ``SPEC'' (``an explicitly declared spectroscopic value''). Note, that was done in \citet{lop18}, so that the NED redshifts are available only for the ESZ+X-ray clusters. Hence, we have only SDSS DR7 redshifts for the NoSOCS clusters and DR14 plus NED redshifts for the ESZ+X-ray sample. That is important to check for possible differences in the membership classification results related to incompleteness in the SDSS survey (see below, $\S$\ref{results}).

\subsection{Spectroscopic Cluster Membership Assignment}
\label{spec_members}

The spectroscopic selection of members and exclusion of interlopers was done in \citet{lop09a, lop14} and \citet{lop18}, for the NoSOCS and ESZ+X-ray samples, respectively. In both cases we employed a method close to the ``shifting gapper'' procedure \citep{fad96}. The main details are given below. For the full description we refer the reader to the works of \citet{lop09a,lop14}. 

For each group, initially we have all galaxies within 2.50 h$^{-1}$ Mpc (3.57 Mpc for $\rm h = 0.7$) of their centers and with a velocity offset of $\lvert \Delta_v \rvert \le 4000$ \kms. The ``shifting gapper'' procedure is based on the application of the gap-technique in radial bins, starting in the cluster center. The bin size is 0.42 h$^{-1}$ Mpc (0.60 Mpc for $\rm h = 0.7$) or larger to force the selection of at least 15 galaxies. Those not associated with the main body of the cluster are eliminated. This procedure is repeated until the number of cluster members is stable (no more galaxies are rejected as interlopers). This method makes no hypotheses about the dynamical status of the cluster. Once we have a member list we obtain estimates of velocity dispersion ($\sigma_P$) and perform a virial analysis, estimating physical radius and mass ($R_{500}$, $R_{200}$, $M_{500}$ and $M_{200}$).

The way we apply this technique has many details, which can be found in \citet{lop09a,lop14}. The most important one is the use of a variable gap (instead of a fixed one), which scales with the number of galaxies in the cluster region and the velocity difference of those belonging to the group or cluster. This velocity difference is defined from the previous radial bin. The variable gap is named {\it density gap} after the work of \citet{ada98}, who introduced the scaling with the number of galaxies. The scaling relative to the velocity difference comes from \citet{lop09a}. The use of the {\it density gap} turns out to be extremely important as it adapts the method to deal with systems spanning a broad richness and size ranges (from groups to massive clusters).

The maximum radius of $\sim$ 4 Mpc is always larger than $R_{200}$ for all clusters in our sample. The same is true for the maximum velocity offset ($\lvert \Delta_v \rvert \le 4000$ \kms). Hence, after running this code we have a list of member galaxies and interlopers within these radial and velocity limits. 

However, for the membership selection from photometric data alone (the goal of this work) we select all galaxies projected along the line of sight of each cluster, not only those within $\lvert \Delta_v \rvert \le 4000$ \kms. Hence, for the photometric membership selection based on a machine learning approach described below we obviously consider all the remaining galaxies along the line of sight as not members. So, for all other galaxies projected within 4 Mpc, but with $\lvert \Delta_v \rvert > 4000$ \kms, we assign a label of interloper.

To summarize, for each cluster we have selected all galaxies from the SDSS main sample (with a magnitude limit of $r_p \sim 18$, where $r_p$ is the Petrosian magnitude in the $r$-band), with $M_r \le M_r^*+3$ and within a radius of 4 Mpc. All galaxies have a spectroscopic classification of member or interloper, which we can use for training and validation purposes. The combined galaxy sample from all clusters comprises 15802 objects, 4420 being within $R_{200}$.

It is important to stress that our spectroscopic membership classification (derived from the ``shifting gapper'' procedure) was previously applied to observational data \citep{lop09a, lop09b, lop14, lop18}. The member selection is robust, leading to reliable estimates of cluster velocity dispersion, physical radius and mass. More recently, our code's performance for cluster mass estimation was compared to several other galaxy-based methods applied to mock catalogs \citep{old15}. The results obtained with our code (labeled SG3 in that paper) were among the top ones, reinforcing its efficiency for membership selection and further estimation of cluster parameters (such as mass).

On what follows, in order to have a fair photometric membership classification experiment we consider at all times only galaxy and cluster parameters derived from photometry. That means we consider, for instance, a photometric redshift ($z_{\text{phot-cl}}$) estimate for the clusters, galaxy absolute magnitudes derived with distance modulus based on their $z_{\text{phot}}$, galaxy local densities obtained with $z_{\text{phot}}$, etc. In other words, besides the spectroscopic membership classification used for training and evaluation, we totally ignore any spectroscopic derived information for our photometric classification procedure. Hence, knowing the actual class from the ``shifting gapper'' technique we use several photometric parameters as input for machine learning algorithms to predict their class. Some of those cluster and galaxy parameters are detailed below.

\subsection{Cluster Photometric Redshift and Physical Radius} 
\label{clsphot}

All NoSOCS clusters had a photometric redshift estimate from DPOSS data \citep{lop04}, which was updated using SDSS data in \citet{lop07}. We adopted the approach outlined in \citet{lop07} to obtain photometric redshift estimates for the ESZ+X-ray sample. Briefly, we compute cluster (background corrected) mean $r$-band magnitudes and median colours (g-i) and (r-i) within a 0.5 $h^{-1}$ Mpc radius. Those colours and magnitude are used in an empirical relation to obtain a cluster photometric redshift estimate \citep{lop07}.

On what regards a photometric estimate of the physical radius ($R_{200}$) we consider a cluster scaling relation derived in \citep{lop09b}. The relation we adopted is between richness in a fixed metric (0.5 $h^{-1}$ Mpc) and $R_{200}$ derived after the virial analysis \citep{lop09a}. The richness estimate is defined as the number of galaxies (background corrected) with $M^*-1 \le M_r \le M^*+2$ in a given aperture, which we consider 0.5 $h^{-1}$ Mpc. Table 11 of \citet{lop09b} list the parameters of the scaling relations of $R_{200}$ to richness in different apertures. 

As the ESZ+X-ray is composed of systems much more massive than those from NoSOCS we decided to derive a new scaling relation between richness and $R_{200}$ for the former sample. We computed richness in the same way as for the NoSOCS sample and obtained a new scaling relation. The "photometric" $R_{200}$ estimates agree with the original one (from the virial analysis) within $40\%$, with most objects showing a fractional difference smaller than $20\%$. The mean relative difference is $-0.07$, with standard deviation of 0.18. The minimum and maximum fractional differences are $-0.38$ and 0.26.

\subsection{Absolute Magnitudes and Colours} 
\label{galphot}

To compute the absolute magnitudes of each galaxy (in $ugriz$ bands) we consider the following formula: $M_x = m_x - D - kcorr - Qz$ ($x$ is one of the five SDSS bands we considered). D is the distance modulus (considering the photometric redshift of each galaxy), $kcorr$ is the k$-$correction and $Qz$ ($Q = -1.4$, \citealt {yee99}) is a mild evolutionary correction. Rest-frame colours are also obtained for all objects. As some galaxies do not have a k$-$correction in the SDSS database, we adopted a k$-$correction that depends on the galaxy observed colour. If the colour is compatible to an early-type colour we apply an elliptical k$-$correction. If it is not we apply a spiral k$-$correction. In \citet{lop09a} we show there is no significant difference in the final results regarding the k$-$correction procedure, especially for low$-z$ objects (as in our work). On comparison to \citet{lop09a}, the main difference in the computation of absolute magnitudes and rest-frame colours is the use of photometric redshift to compute the distance modulus and for the evolutionary correction. As explained above, we proceed this way as we want to ignore the spectroscopic information when deriving galaxy parameters to be used in the photometric membership classification developed here.

\subsection{Photometric Local Galaxy Density} 
\label{galrho}

We consider the parameter $\Sigma_5$ as our local galaxy density estimator. As in \citep{lop14}, for every galaxy member we compute its projected distance, d$_5$, to the 5th nearest galaxy found around it. The local  density $\Sigma_5$  is simply given by 5/$\pi$d$_N^{2}$, and is measured in  units of galaxies/Mpc$^2$. An important point to keep in mind is that, although our sample has galaxies down to $M^* + 3.0$, the local density we measure is of bright galaxies. Hence, the 5th nearest neighbor is searched from a list of galaxies brighter than $M^* + 1.0$. The main difference for a spectroscopic density estimate is that we restricted the neighbor search to galaxies with a maximum velocity offset of 12000 \kms, instead of 1000 \kms. That takes in account the photometric redshift uncertainty ($\Delta_z \sim 0.02$). We tested different velocity cuts to minimize the background and found the above value to be reasonable for our purpose, which is to differentiate members and interlopers according to their local density distributions (see Fig.~\ref{fig:dist_6pars} below). 

\section{Machine Learning Methods}

In almost all science fields we can find applications of machine learning (ML) techniques. ML algorithms can learn from data, improving from experience, not requiring human intervention. They can also be understood as methods to learn a function that best maps input variables to an output.

We can separate ML algorithms according to their learning styles. They are termed methods of {\it supervised} and {\it unsupervised} learning. In the first case, we know a label or result of the data, which we want to be able to predict for any other data set after a training process. Classification and regression are common problems tackled by supervised learning methods. In the unsupervised case the input data has no previously known label or result. The model works by deducing structures present in the input data. Dimensionality reduction and clustering are typical problems targeted by unsupervised methods. There is also the semi-supervised case for which the input data is a mixture of labeled and unlabelled sets. 

Algorithms can also be grouped by similarity on how they work, instead of their learning styles. The most common groups are called {\it regression}, {\it instance-based}, {\it regularization}, {\it decision tree}, {\it bayesian}, {\it clustering}, {\it artificial neural network }, {\it dimensionality reduction}, and {\it ensemble} algorithms\footnote{it is possible to find a larger list and a detailed explanation of each algorithm at https://machinelearningmastery.com/a-tour-of-machine-learning-algorithms/}. 

Many ML applications in astronomy are found in the literature. A recent review is presented by \citet{bar19}. A few examples are the estimation of photometric redshifts \citep{bal07,car13}, galaxy morphological classification \citep{hue08,hoc18}, star/galaxy separation \citep{ode04,kim15}, membership assignment in star clusters \citep{kro14}, and transit detection \citep{san19}.

There are only a few studies that have addressed the issue of galaxy membership assignment in galaxy clusters from photometric information and none of them rely on ML methods \citep{geo11,bru00,roz15,cas16,bel18}. The usual approach is based on calculating membership probabilities from the photometric redshift distributions and/or assuming typical spatial and luminosity profiles for galaxies in clusters. To the best of our knowledge the current work is the first to address this issue with a machine learning method.

\subsection{The Caret R Package}

The machine learning approach we adopt makes use of the Caret\footnote{https://topepo.github.io/caret/}$^{,}$\footnote{ https://cran.r-project.org/web/packages/caret/} package, created and maintained by Max Kuhn, and available within the computer language R. Caret stands for {\it Classification And REgression Training}. This package comprises a set of functions that attempt to facilitate the process of creating predictive models. At first, the package was designed to unify model training and prediction. Later, it expanded to standardize common tasks such as parameter tuning and estimating variable importance (see $\S$\ref{featsel}). The Caret package contains many tools for different purposes, such as data splitting and pre-processing, feature selection and importance, parallel processing and visualization.

Although many of the model functions (if not all) we use in this work are already available in R, we opted for using Caret due to its uniformity, consistency and easiness for working with machine learning algorithms. 

\subsection{Membership Probabilities}
\label{memb_prob}

All the machine learning models we employ below provide for each galaxy the probability that it belongs to a cluster, which we call the membership probability ($P_{\text{mem}}$). A galaxy is classified as a member if it has a $P_{\text{mem}}$ value higher than a given probability threshold ($P_{\text{th}}$). The results we show below consider the default probability threshold $P_{\text{th}} =$ 0.5 (a galaxy should have more than 50$\%$ of chance to belong to a cluster).

\begin{figure}
\begin{center}
\leavevmode
\includegraphics[width=3.5in]{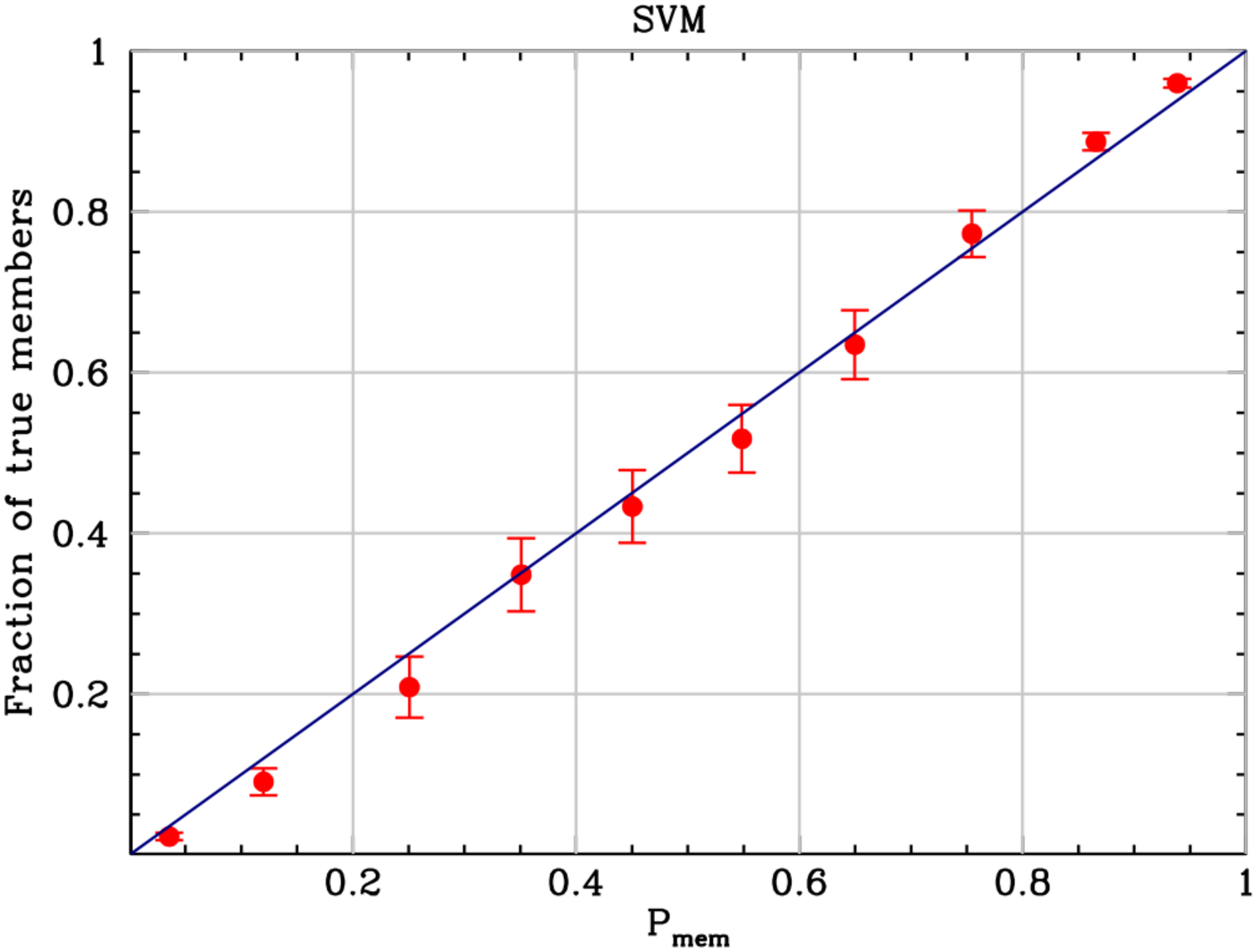}
\end{center}
\caption{Fraction of true members as a function of the membership probability ($P_{\text{mem}}$).}
\label{fig:fract_pmemb}
\end{figure}

In order to check the significance of the membership probabilities we can compare the fraction of true members ($f_{\text{true}}$) to $P_{\text{mem}}$. As in \citet{cas16} we binned the data in $P_{\text{mem}}$, computing the fraction of true members for each $P_{\text{mem}}$ bin. That is shown in Fig.~\ref{fig:fract_pmemb}. We found a good agreement between $f_{\text{true}}$ and $P_{\text{mem}}$, reassuring the quality of the probabilities we derive. The mean difference and the rms dispersion around the mean are $\langle f_{\text{true}} - P_{\text{mem}} \rangle =$ -0.0087 $\pm$ 0.0230.

Although we adopted the default probability threshold ($P_{\text{th}} =$ 0.5) we still discuss how our results would change for different choices of $P_{\text{th}}$ (see $\S$\ref{prob_thre}).

\subsection{Completeness and Purity}
\label{comp_pur}

Any classification experiment results in false positives (a positive result when it should be negative) and false negatives (negative output when it should be positive). That is also the case for the galaxy membership assignment problem of this work. We quantify the robustness of our photometric membership classification in terms of completeness ($\text{C}$) and purity ($\text{P}$). As described in \citet{geo11} completeness and purity are measurements of the relation between the sample of objects classified as members and the true population of members. Completeness is also known as the "True Positive Rate" (TPR) or "Sensitivity", while purity is known as "Precision" or "Positive Predictive Value" (PPV).

Completeness is defined as the fraction of true members that are  classified as members, while purity is the fraction of true members among the objects classified as members. Following the notation of \citet{geo11} and \citet{cas16} we call $N_{\text{selected}}$ the number of galaxies that are photometrically classified as members, while $N_{\text{true}}$ is the true number of members, $N_{\text{interlopers}}$ is the number of objects wrongly classified as members, and $N_{\text{missed}}$ is the number of true members that were not photometrically classified as so. $N_{\text{interlopers}}$ are the false positives, while $N_{\text{missed}}$ are the false negative results. According to those definitions we have $N_{\text{true}} = N_{\text{selected}} + N_{\text{missed}} - N_{\text{interlopers}}$ and

\bea
\text{C} = \frac{N_{\text{true}} - N_{\text{missed}}}{N_{\text{true}}}
\label{eq:purity}
\eea

\bea
\text{P} = \frac{N_{\text{selected}} - N_{\text{interlopers}}}{N_{\text{selected}}}
\label{eq:completeness}
\eea

\bea
N_{\text{true}} = \frac{\text{P}}{\text{C}} N_{\text{selected}}.
\label{eq:ntrue}
\eea

The last equation (\ref{eq:ntrue}) provides a powerful way to recover the true number of galaxies (richness of a cluster) from the estimated value.

To properly assess the performance of our photometric membership classification we divide our sample in two, labeled the training and validation data sets. Hence, we train the algorithms in the first set (training) and test their performance in the second (validation). It is common practice in the literature to split the data in $80\%$ for training and $20\%$ for validation. Instead, we split in half. We do so as some of the analysis we show below further split the data in magnitude, colour, clustercentric distance and parent cluster mass. Hence, in order to have a reasonable number of galaxies on each bin we decided to have $50\%$ of the data in the validation set. That gives 2210 galaxies, as the total number for the combined NoSOCS and ESZ+X-ray samples is 4420, with $M_r \le M_r^*+3$ and within $R_{200}$ ($\S$\ref{spec_members}). 

\subsection{Feature Selection}
\label{featsel}

Having already built a data set in $\S$\ref{data} the first task we face is the feature selection. There are different ways to test and select the most relevant features for a given problem. One can search for correlations between different attributes and remove those that are highly correlated. Another option is to estimate the importance of features when building a model. The concept of "importance" helps us to explain the predictive power of the features in the dataset through a "score". The higher the score the more important or relevant is the feature. We used a {\it Learning Vector Quantization} (LVQ) model to find the scores associated with the features in our data. Basically, LVQ is a technique of dimension reduction. We initially have an input vector consisting of the training set with observations, each one associated with a set of classes. A class is a template for defining objects, describing what features the object will have. For each class a subset of vectors (taken from the input vector) is assigned, building a number of "reconstruction" vectors. It is assumed that a given observation belongs to the same class to which the nearest reconstruction vector belongs. The LVQ algorithm minimizes the difference between the input vector and the reconstruction vectors through the nearest neighbor rule, using an Euclidean metric, which provides the score for each feature in the dataset. In short, the LVQ algorithm is a type of artificial neural networks that allows one to choose the number of training features in which the essential information content of the input data is concentrated (see \citealt{koh95} for details).

\begin{figure}
\begin{center}
\leavevmode
\includegraphics[width=3.5in]{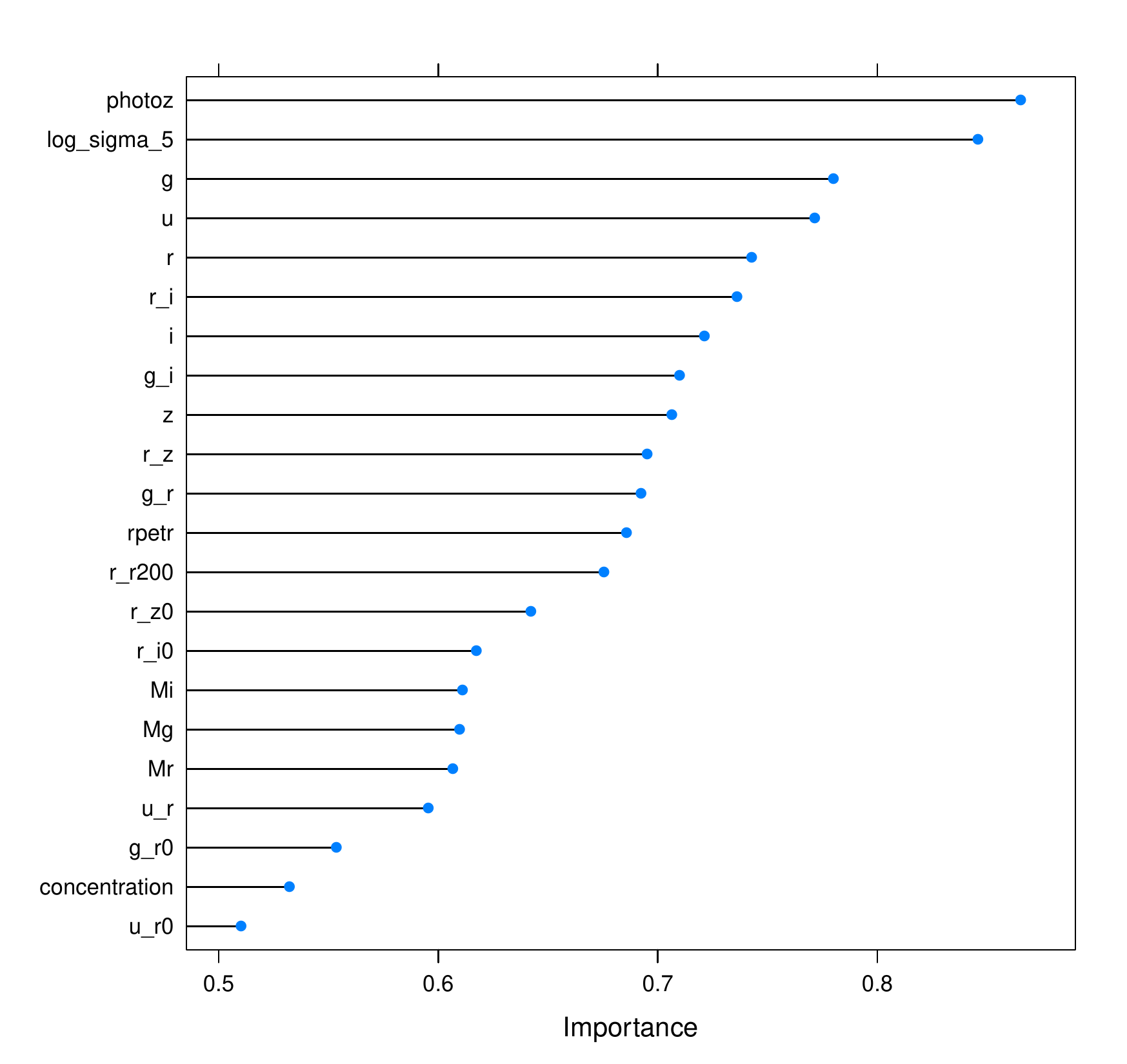}
\end{center}
\caption{Importance of different variables available in the data set. From top to bottom, the variables listed are $z_{\text{phot}}$, LOG $\Sigma_5$, g, u, r, (r-i), i, (g-i), z, (r-z), (g-r), $R_{\text{petr}}$, $R/R_{200}$, (r-z)$_0$, (r-i)$_0$, M$_i$, M$_g$, M$_r$, (u-r), (g-r)$_0$, concentration ($C$) and (u-r)$_0$.}
\label{fig:importance_feat}
\end{figure}

In Fig.~\ref{fig:importance_feat} we can see the features ranked by importance, with the $z_{\text{phot}}$, local density, observed magnitudes and colors among the most important. A few other methods, such as {\it Support Vector Machines}, also give similar results. 

\begin{figure}
\begin{center}
\leavevmode
\includegraphics[width=3.5in]{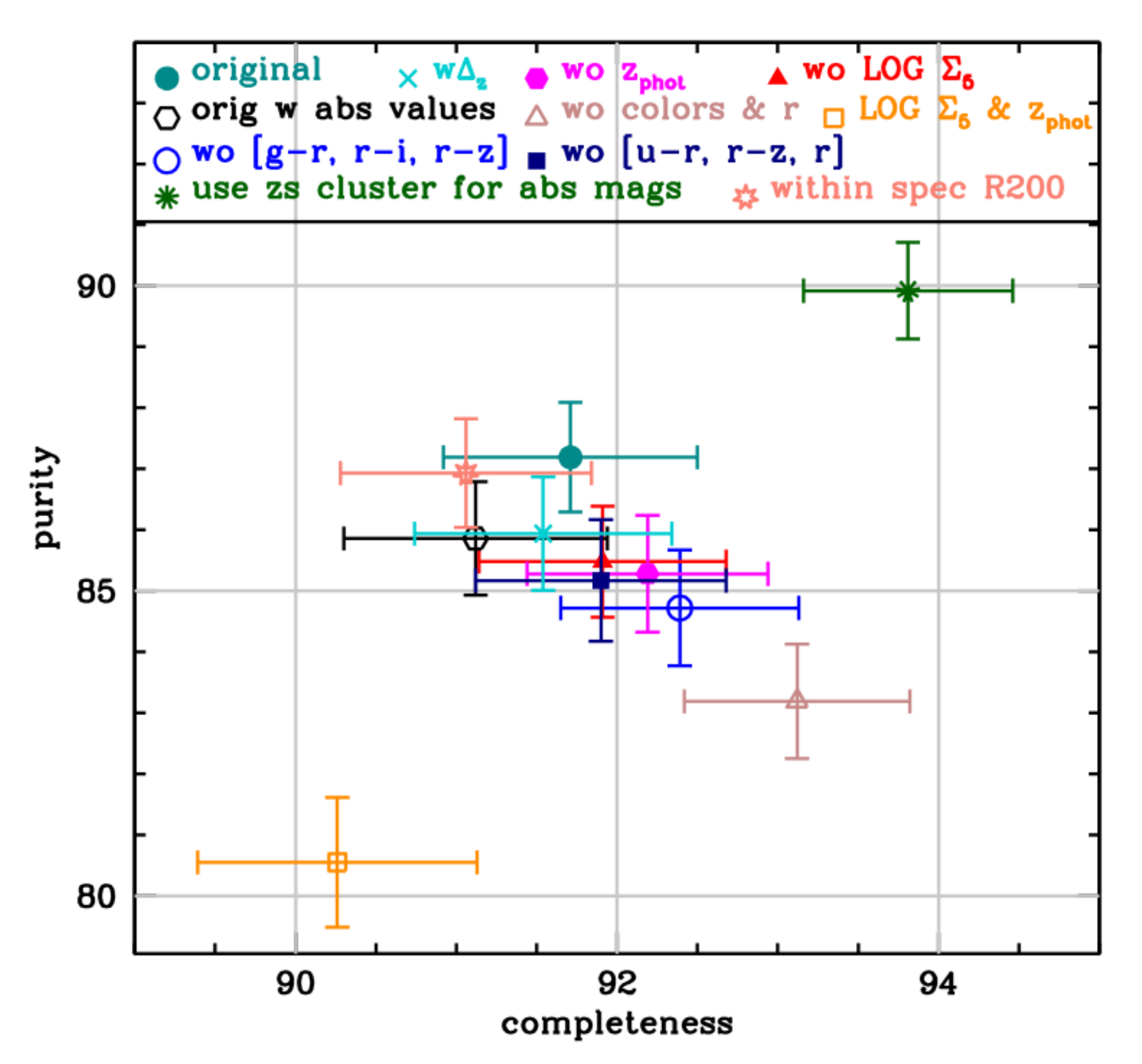}
\end{center}
\caption{Purity {\it vs} completeness obtained with the SVM algorithm (for the validation sample) with different sets of input variables. Each point represents the result derived with different features. Our final selection (called original in the figure) is represented by the dark cyan filled circle, which has the result based on (u-r), (g-r), (g-i), (r-i), (r-z), r, LOG $\Sigma_5$, $R_{\text{petr}}$, $C$, $R/R_{200}$ and $z_{\text{phot}}$; the dark turquoise cross is based on the same configuration, but uses (for each galaxy) the offset between its $z_{\text{phot}}$ and the cluster $z_{\text{phot-cl}}$, instead of the galaxy $z_{\text{phot}}$; the filled magenta hexagon is also as the original configuration, but without $z_{\text{phot}}$; the red filled triangle does not use LOG $\Sigma_5$, instead; the black open hexagon is analogous to the first configuration (dark cyan filled circle), but with rest-frame photometric parameters (for magnitude and colors), instead of the observed ones; the light brown open triangle has the same configuration as the original (dark cyan filled circle), but without all the colors and magnitude; the dark orange open square is for the case where we considered only the top two features (LOG $\Sigma_5$ and $z_{\text{phot}}$); the open blue circle is analogous to the original, but without three colors (g-r, r-i, r-z); the navy blue filled square is also analogous to the original, but without (u-r), (r-z) and r. The last two points are shown for different purposes. The salmon open star shows the results obtained when considering the spectroscopically derived $R/R_{200}$ (instead of the photometric). The green asterisk shows the improvement in the results if we knew the spectroscopic redshift of the clusters and simply consider those for computing absolute magnitudes. That affects the final galaxy selection ($M_r \le M^*+3$), as well as the density estimates.}
\label{fig:comp_pur_inp_var}
\end{figure}

Nonetheless, we notice that this is not always the case, as some 
methods employed to check the feature importance could lead to a different feature selection. For instance, assessing the best features with the RF model we find the top eight include $z_{\text{phot}}$ and local density (as above), but also the Petrosian radius in the $r$-band ($R_{\text{petr}}$), galaxy concentration ($C$) and normalized radial distance to the cluster center ($R/R_{200}$), as well as three colors. Hence, we took the results of the feature selection as a initial guess, but tested the performance of our algorithms with different variables. That is seen in Fig.~\ref{fig:comp_pur_inp_var}, which displays purity {\it vs} completeness obtained with the SVM algorithm (the definitions of purity and completeness are found in $\S$\ref{comp_pur}). All results shown in Fig.~\ref{fig:comp_pur_inp_var} are derived with the validation sample\footnote{The choice for 50\% is explained in $\S$~\ref{comp_pur}.} (50\% of our original data). In almost all cases, each point in the figure shows the results derived when using different sets of input variables. The exception is for two situations, for which we test the impact of assuming the spectroscopically derived $R_{200}$ or using the clusters' spectroscopic redshifts to estimate absolute magnitudes. In most of the other cases, based on different input variables, this figure indicates no significant difference between the results.

\begin{figure*}
\begin{center}
\leavevmode
\includegraphics[width=6.5in]{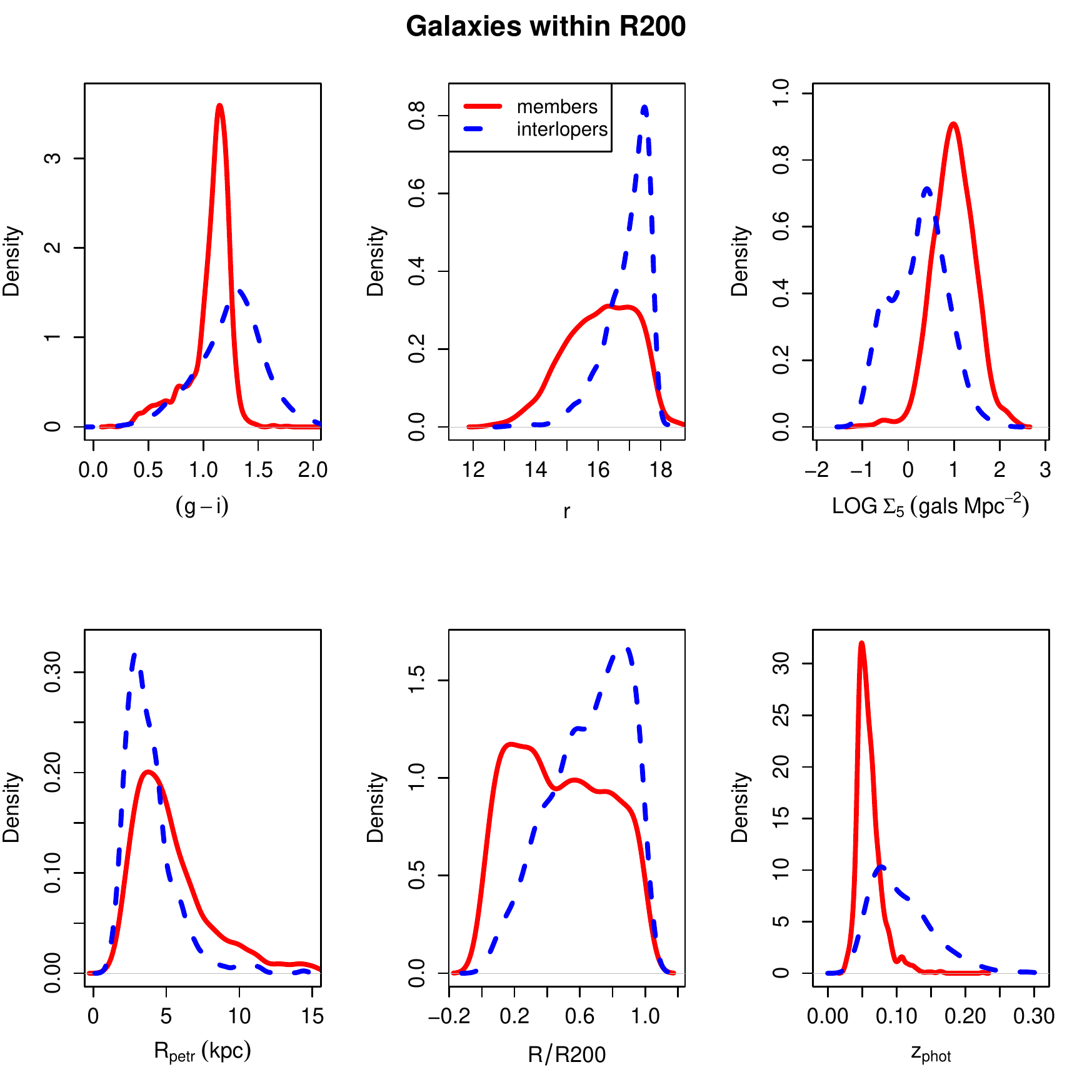}
\end{center}
\caption{Distribution of six photometric parameters of members (red) and interlopers (blue) of the ESZ+X-ray sample. The membership classification considered for this figure is based on the spectroscopic data (see $\S$\ref{spec_members})}.
\label{fig:dist_6pars}
\end{figure*}

After those tests we decided to work with the following set of parameters: (u-r), (g-r), (g-i), (r-i), (r-z), r, LOG $\Sigma_5$, $R_{\text{petr}}$, $C$, $R/R_{200}$ and $z_{\text{phot}}$ (result displayed by the dark cyan filled circle in Fig.~\ref{fig:comp_pur_inp_var}). As indicated above, all the remaining results (except for two) of Fig.~\ref{fig:comp_pur_inp_var} are variations from this configuration. We show in Fig.~\ref{fig:dist_6pars} the distribution within $R_{200}$ of members and interlopers (spectroscopic classification) of six parameters of the galaxies in the ESZ+X-ray sample. As it can be seen, members (red) and interlopers (blue) have distinct distributions, which reinforces that those photometric parameters can indeed be used for membership classification. Even a parameter such as $R_{\text{petr}}$ that has a large overlap between the distributions of members and interlopers is useful for the machine learning models. For the current data (limited to clusters with $z_{\text{phot-cl}} \le 0.045$ and galaxies with $r \la 18$) we see the interlopers are generally fainter than the members, also showing significant redder colours.

It is also important to notice in Fig.~\ref{fig:comp_pur_inp_var} that the use of the original $R_{200}$ value (based on the galaxy velocities) does not lead to more accurate results (compare the salmon open star to the original results given by the dark cyan filled circle). On the contrary, the knowledge of the clusters' spectroscopic redshifts can significantly improve the classification (see the green asterisk).

\subsection{Model Selection and Tune}

After choosing the set of features we tested the performance of eighteen different machine learning algorithms. We selected the six algorithms with better performance: {\it Support Vector Machines with Radial Basis Function Kernel}, {\it Stochastic Gradient Boosting}, {\it Model Averaged Neural Network}, {\it knn}, {\it Random Forest}, {\it C5.0}.

\begin{figure}
\begin{center}
\leavevmode
\includegraphics[width=3.5in]{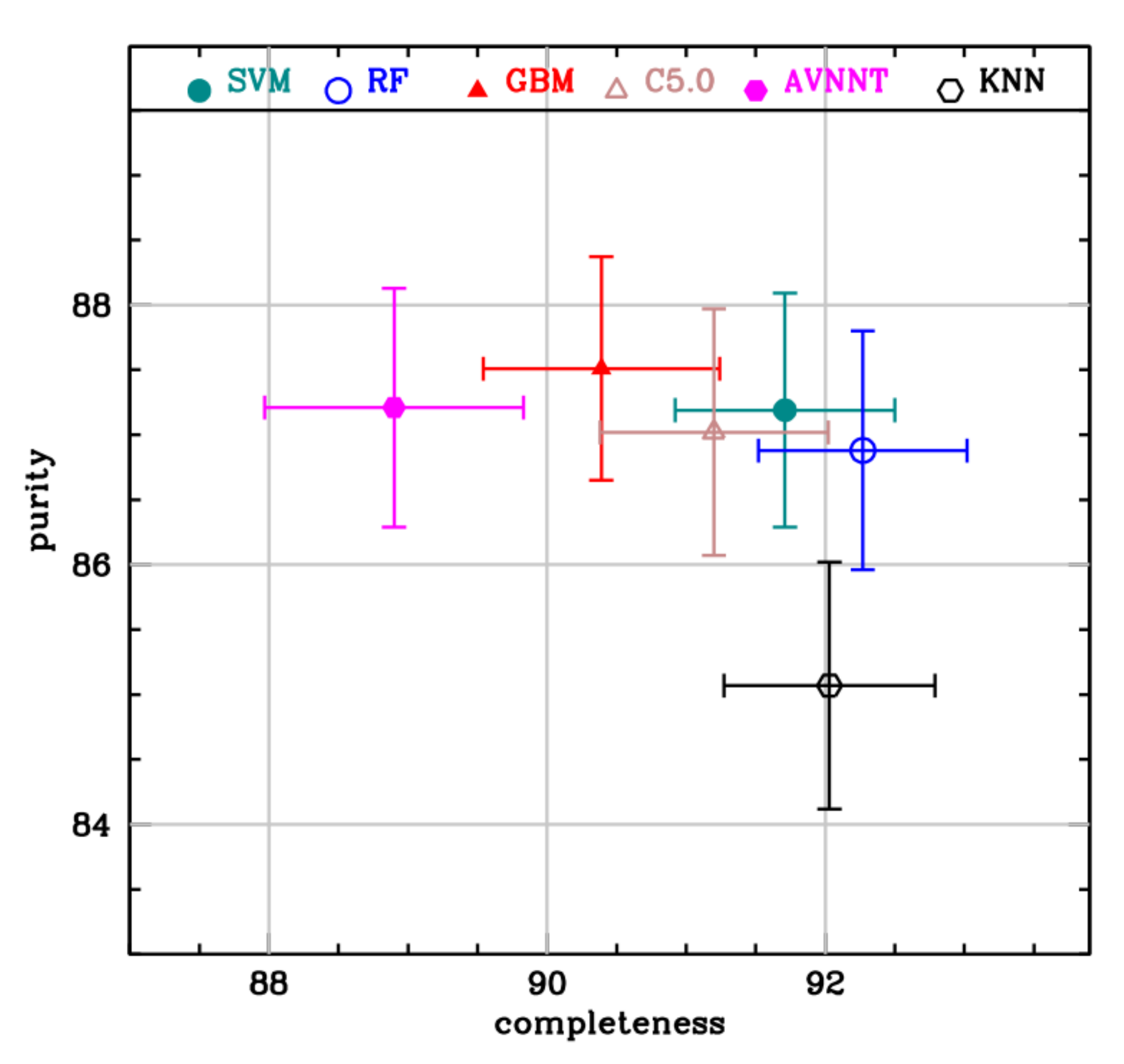}
\end{center}
\caption{Purity {\it vs} completeness obtained with the the six best models tested (for the validation sample).}
\label{fig:comp_six_alg}
\end{figure}

Next, as the default configuration of each algorithm may not be the best, we tune them searching for the best parameter configuration within each model. After selecting the best configuration for each algorithm we compared their performance one more time. That can be seen in Fig.~\ref{fig:comp_six_alg} where we show the purity {\it vs} completeness obtained for those six models. Although the results are very similar we consider our final results as those given by the {\it Support Vector Machines with Radial Basis Function Kernel} (simply called SVM) model. However, it is important to note the conclusions could be different for other data sets. A brief description of the SVM algorithm is given below.

\subsubsection{Support Vector Machines with Radial Basis Function Kernel}

Support vector machines (SVM) is a supervised learning algorithm that can be used for classification and regression problems. The basic idea behind SVM is the search of a hyperplane best separating the features into different domains. The optimal hyperplane is the plane with the maximum distance between the plane and the data points. Once the hyperplane is found new objects are classified according to their distance to the hyperplane \citep{bar19}. The points that are closer to the hyperplane are named the support vector points, while their distance from the hyperplane are called the margins.

It is common to find problems where a linear separation is not possible for the classes. In that case, a {\it kernel trick} can be employed to transform the dataset into a higher dimension feature space, where the classes might be linearly separated. After a decision boundary is derived it can be projected back to the original input space. A large variety of kernel functions can be employed, such as linear, polynomial and the radial basis function kernel (RBF). The RBF kernel is a function that depends on the distance from the origin or from some other point. 

A main advantage of SVM is its effectiveness for higher dimension classification problems. The method is very sensitive to the measured distances between the objects and the hyperplane, but that can be overcome through feature scaling.

\section{Results}
\label{results}

In this work we photometrically selected members within $R_{200}$. For the two samples we considered we got the following results with the SVM model. For NoSOCS, we found C $ = 93.2\% \pm 1.0\%$ and P $ = 84.7\% \pm 1.4\%$, while for the ESZ+X-ray sample we have C $ = 92.7\% \pm 0.9\%$ and P $ = 86.8\% \pm 1.1\%$. The small difference in purity may be due to incompleteness in the spectroscopic sample used for NoSOCS (based in the SDSS DR7). The ESZ+X-ray has SDSS data complemented with NED redshifts. As shown in \citet{lop18} the inclusion of NED redshifts helps alleviating the incompleteness in the SDSS main sample, resulting in a better separation of members and interlopers. The under sampling of spectroscopic coverage, even for a survey like the SDSS, is an important issue. For instance, \citet{yoo08} mention a 30-40\% incompleteness rate in dense regions of the SDSS, when selecting galaxy clusters. \citet{von07} states that the central galaxy has no spectroscopic redshift  $\sim$30\% of times in SDSS. \cite{spe16} address this issue investigating different counting techniques to overcome incompleteness in the spectroscopic sample of SDSS.

The results derived with the combined data set (NoSOCS and ESZ+X-ray samples) described in $\S$\ref{data} are C $ = 91.7\% \pm 0.8\%$ and P $ = 87.2\% \pm 0.9\%$. These data contains 4420 galaxies within $R_{200}$. Except for the results described above (based only on NoSOCS or the ESZ+X-ray sample) and the discussion regarding system mass below, all the other results presented in the current work are derived from this combined sample. 

\subsection{Variation with magnitude and colour}

\begin{figure}
\begin{center}
\leavevmode
\includegraphics[width=2.5in,angle=270]{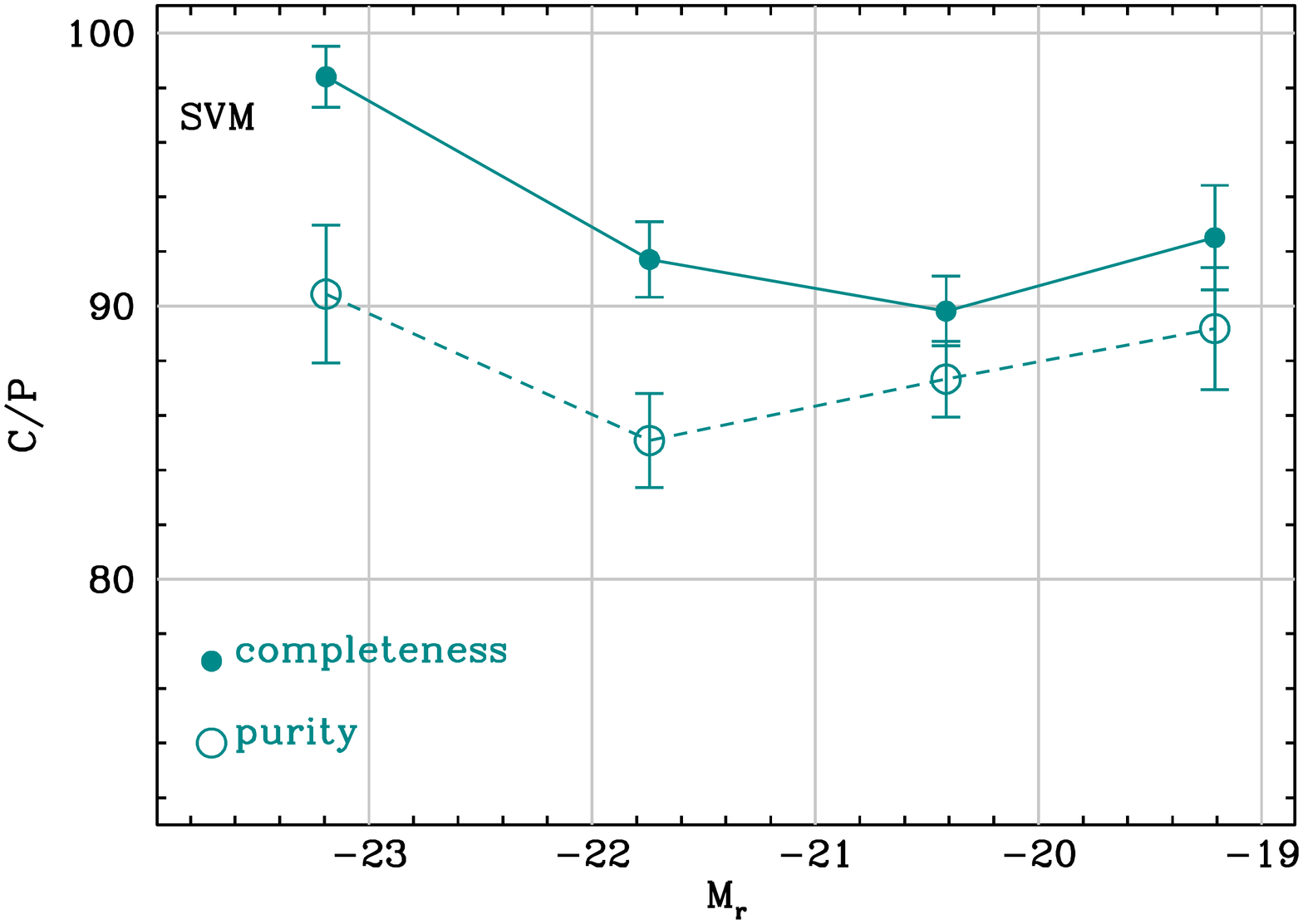}
\end{center}
\caption{Variation of completeness (C) and purity (P) as a function of absolute magnitude in the $r-$band for the combined NoSOCS and ESZ+X-ray sample. C is shown with filled symbols and solid curve, while P is displayed with open points and dashed lines. The error bars indicate the standard error of a proportion.}
\label{fig:comp_pur_magr}
\end{figure}

\begin{figure}
\begin{center}
\leavevmode
\includegraphics[width=2.5in,angle=270]{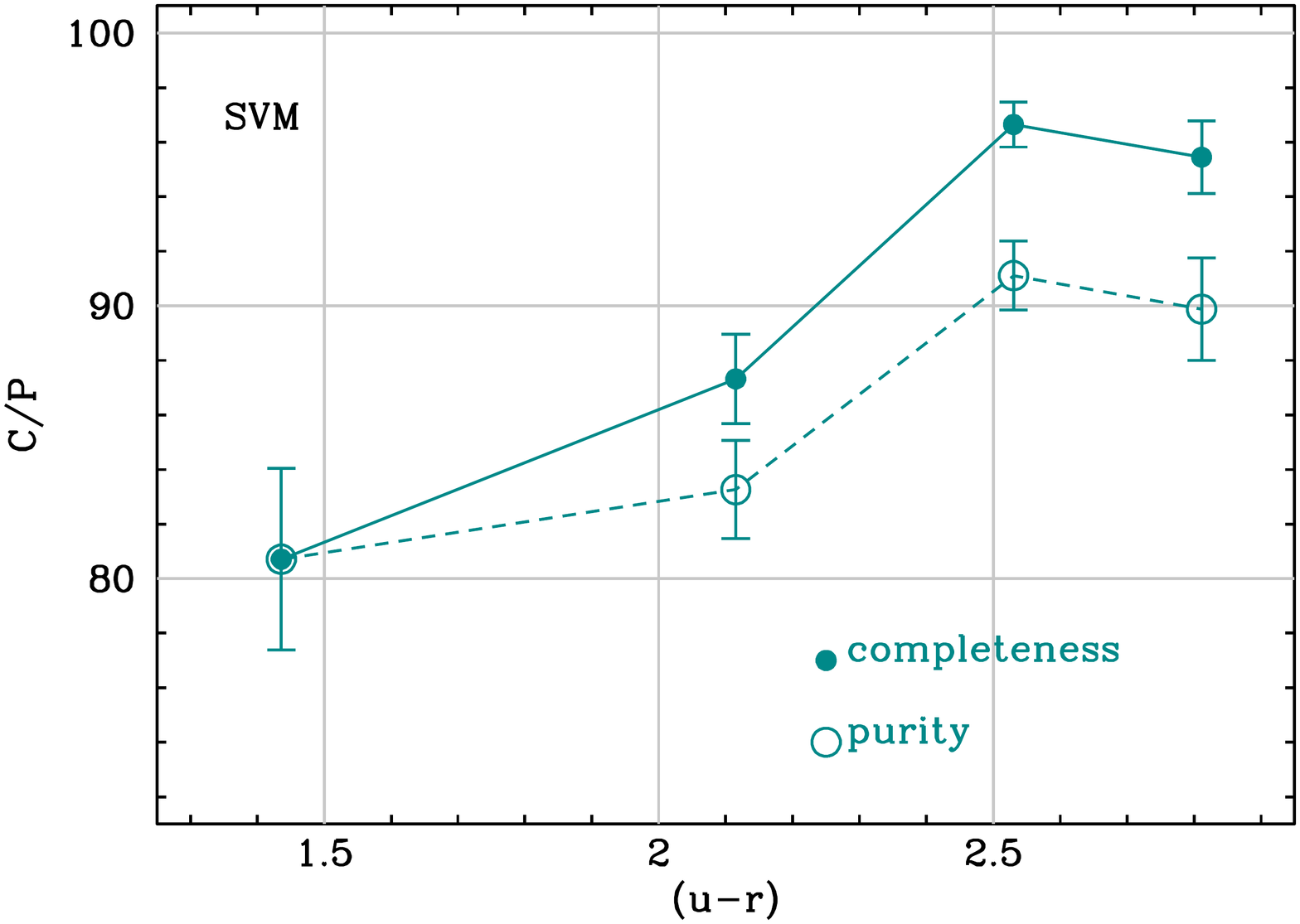}
\end{center}
\caption{Same as in the previous figure, but showing the dependence with the (u-r) colour for the combined NoSOCS and ESZ+X-ray sample. The error bars indicate the standard error of a proportion.}
\label{fig:comp_pur_colour}
\end{figure}

Next, we investigate the possible dependence of the classifier's performance on magnitude and colour. That is shown in Figs. \ref{fig:comp_pur_magr} and \ref{fig:comp_pur_colour}, displaying the variation of completeness (C) and purity (P) according to absolute magnitude in the $r-$band and (u-r) colour, respectively. Those two figures show the results for the combined data set (NoSOCS and ESZ+X-ray samples). We notice that completeness is generally above 90\%, while purity most of times is above 85\%. 

On what regards magnitude (Fig.~\ref{fig:comp_pur_magr}) we see just a small variation of C and P, from the first bin (with brighter galaxies) to the second one. C drops from $\sim$ 98\% to $\sim$ 92\%, remaining nearly constant for lower luminosities. A similar behaviour is seen for purity (around 87\%). That is reassuring, indicating the classification works well, no matter the magnitude, down to $M^* + 3.0$ ($M_r^* \sim -21.6$).

The plot of C and P against colour (Fig.~\ref{fig:comp_pur_colour}), display a performance improvement, from blue to redder colors, as expected, as the red galaxies are dominant within clusters. C and P are about 80\% for galaxies bluer than $(u-r) = 1.5$, increasing to C $\sim$ 95\% and P $\sim$ 90\% for $(u-r) > 2.5$. One could expect the results to be better than what we found for blue galaxies, as those objects could have distinct properties according to environment. In other words, perhaps the separation of blue field and cluster galaxies could lead to higher C and P values. In that case, there could be a small decrease in the efficiency of the classifier in the transition region \citep{str01, lop14, lop16} between blue and red galaxies, with better results in the extremes. We are not sure that could be the case for our classification. Perhaps, the small sample size (30 clusters) and the low redshift limit ($z_{\text{phot-cl}} = 0.045$), prevents us reaching clearer results. Nonetheless, it is still reassuring we can photometrically select blue cloud galaxies in clusters with C and P $\gtrsim$ 80\%.

\subsection{Dependency on cluster mass}

\begin{figure}
\begin{center}
\leavevmode
\includegraphics[width=2.5in,angle=270]{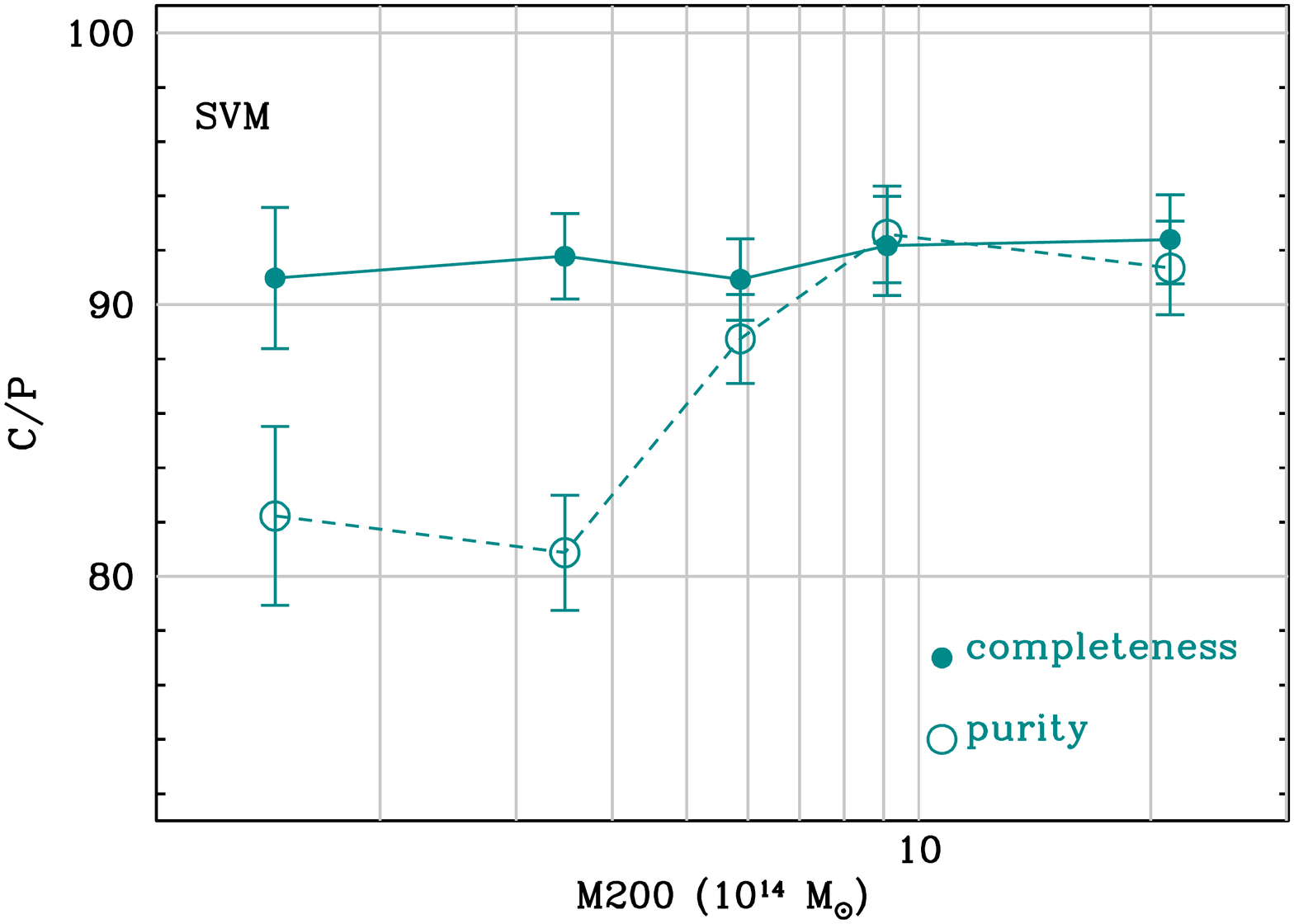}
\end{center}
\caption{Same as in the previous figure, but exhibiting the variation with parent cluster mass for the combined NoSOCS and  ESZ+X-ray sample. The error bars indicate the standard error of a proportion.}
\label{fig:comp_pur_clsmass}
\end{figure}

We have also investigated if the results could depend on the parent cluster mass. However, Fig.~\ref{fig:comp_pur_clsmass} shows no significant variation of completeness relative to cluster mass. On the contrary, P is approximately constant ($\sim$ 82\%) for lower masses ($M_{200} \la 4 \times 10^{14} M_{\odot}$), increases to $\ga$ 90\% at $M_{200} \sim 6 \times 10^{14} M_{\odot}$, remaining constant for higher masses. Although the lower mass limit of our sample is $0.59 \times 10^{14} M_{\odot}$, we are biased towards more massive clusters (the mean mass of the sample is $5.05 \times 10^{14} M_{\odot}$). In order to check if the results get worse for lower masses we make use of 241 groups and clusters from the FoF group catalog of \citet{ber06}. The original catalog was built based on the SDSS DR3. We actually use this catalog's extension to DR7 \citep{lab10,rib13a}. Those 241 systems have $z_{\text{phot-cl}} \le 0.045$ and had their galaxy properties selected as for the NoSOCS and ESZ+X-ray samples. The photometric membership classification based solely on the 241 systems from \citet{ber06} results in a loss of efficiency, with C and P $\sim$ 75\%. It is important to stress this latter sample \citep{ber06} was selected in a very different way from the other two. These lower mass systems could be more affected by a number of biases, such as miscentering. The small number of galaxies per system may also lead to more mistakes in the original spectroscopic membership assigment. Due to that and the fact the main goal of the current work is to investigate the photometric membership selection in clusters, we defer a deeper analysis regarding the system mass to a future work. We plan to use groups and clusters from cosmological simulations, for which the halo center is properly defined and spectroscopic incompleteness is not a problem.

\subsection{Choice of probability threshold} 
\label{prob_thre}

A major issue in classification problems is present when classes have a severe imbalance. As a consequence, the performance could result very biased against the class with smallest frequencies. In our case, if we had, for instance, a much larger fraction of members than interlopers in our spectroscopic classes, the predictive models could maximize accuracy by predicting all galaxies to be members. We would have great results on completeness, but at the cost of low purity.

\begin{figure}
\begin{center}
\leavevmode
\includegraphics[width=3.5in]{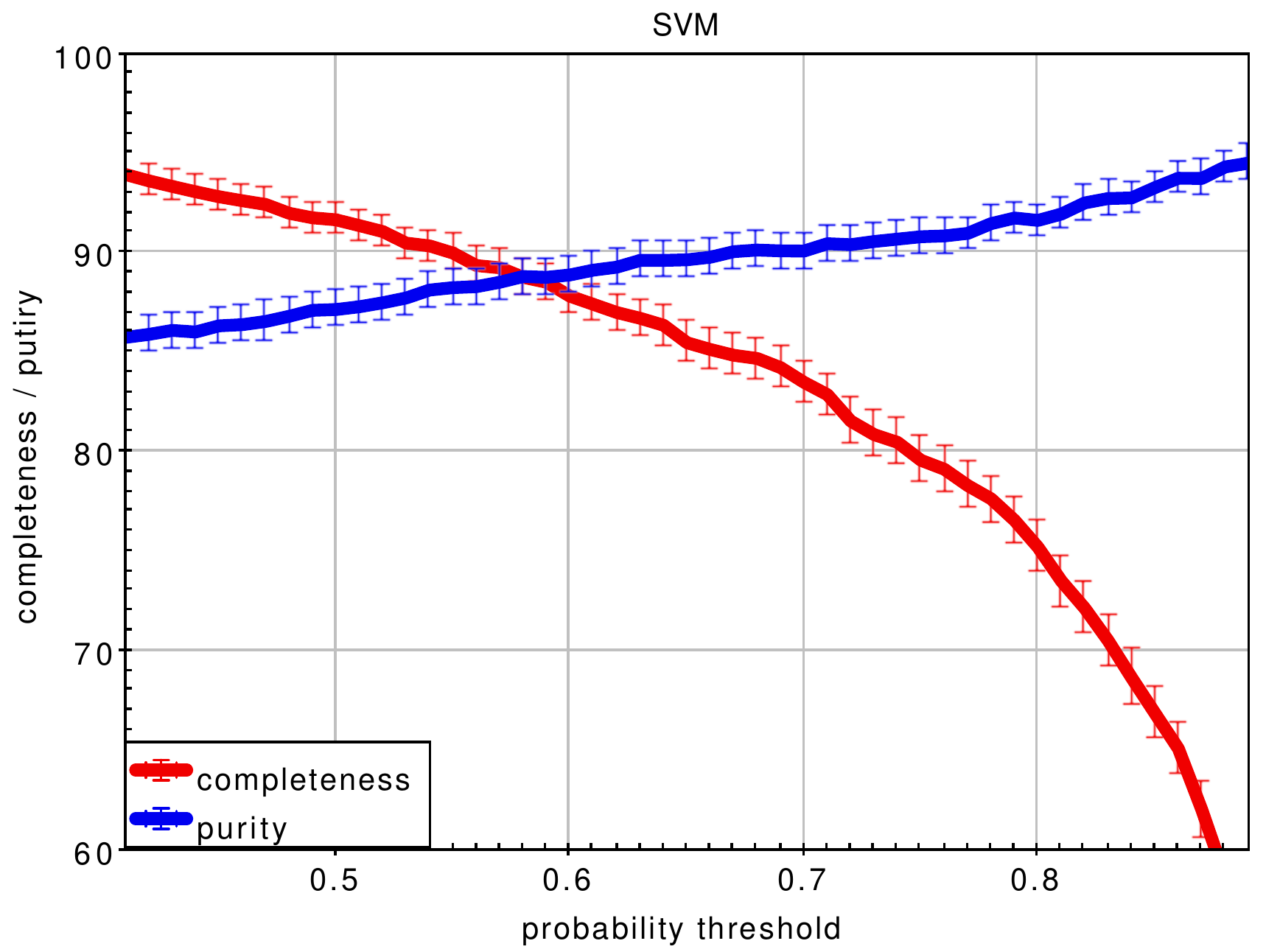}
\end{center}
\caption{The variation of completeness (red) and purity (blue) as a function of the probability threshold, for the SVM algorithm.}
\label{fig:comp_pur_thre}
\end{figure}

A common approach to this problem relies on different subsampling techniques, like down and up-sampling. Another possibility is to use the ROC curve to search for an alternative probability cut off. By default the "Positive class" classification is done at a probability cut off value of 50\%. In our case, galaxies are called members if they have a probability greater or equal to 50\% of being so. However, that might not be the best choice. Instead of using the ROC curve to search the best threshold we inspect the variation of C and P as a function of the probability threshold. In Fig.~\ref{fig:comp_pur_thre} we see the expected behavior of decreasing completeness and increasing purity as a function of the probability threshold. For low probability cuts ($< 0.20$) C is larger than 99\%, but purity is $\sim$ 75\%. The optimal threshold would be $\sim 0.57$, where we would have nearly equal completeness and purity (C $=$ P $\sim$ 89\%). The default threshold cut (0.50) results in C $=$ 91.7\% and P $=$ 87.2\%. As the increase in P is small we decided to keep our main results as obtained with the default probability cut. However, different users could opt for a new cut, depending on their goals. In any case, as the probabilities are available they can also be used to compute cluster richness. Instead of simply counting the number of members classified at a given probability threshold, one could compute richness as the sum of all galaxies' membership probabilities \citep{geo11}.

\begin{figure}
\begin{center}
\leavevmode
\includegraphics[width=2.5in,angle=270]{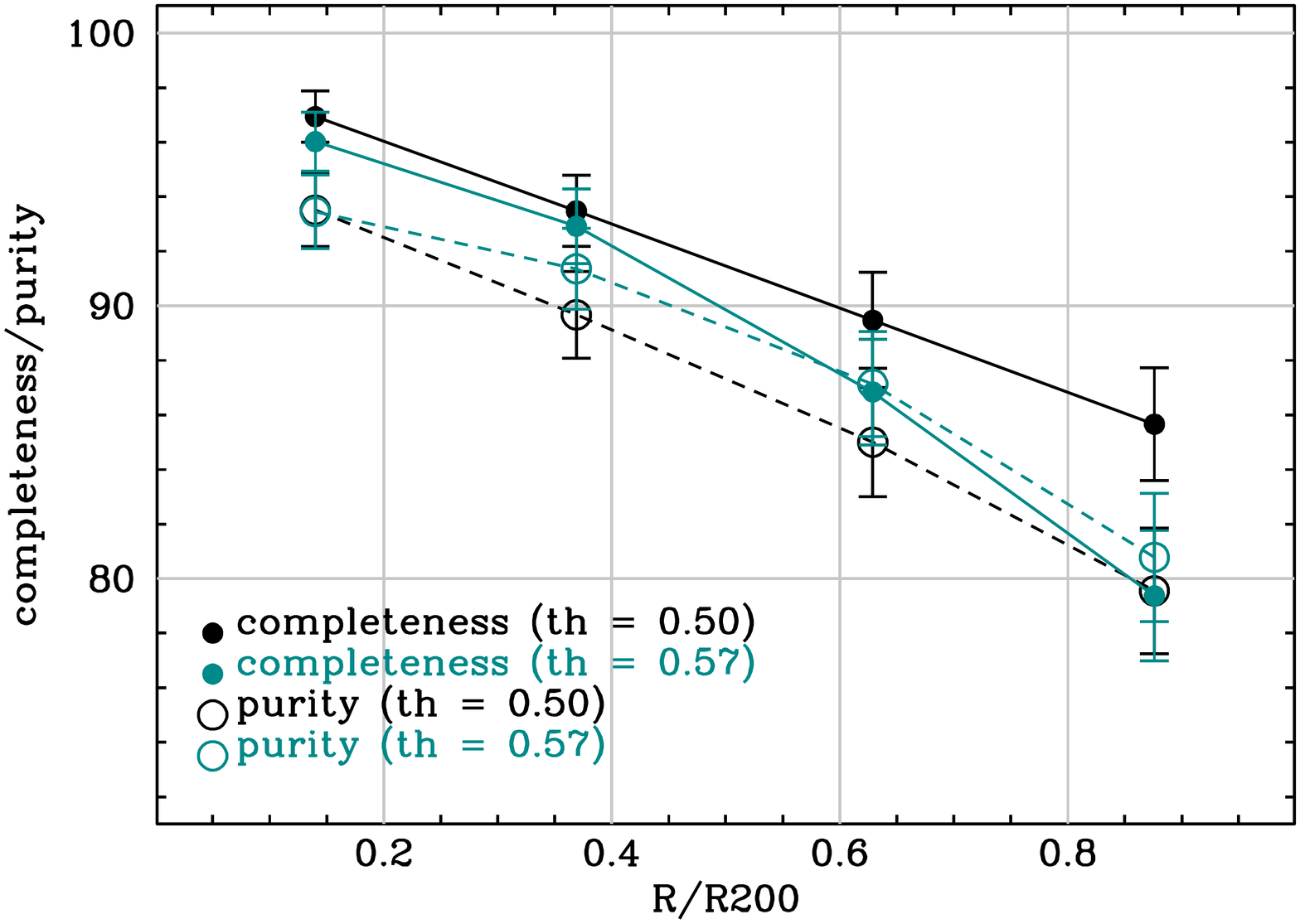}
\end{center}
\caption{The variation of completeness (solid lines and filled symbols) and purity (dashed lines and open symbols) as a function of the normalized distance to the cluster center. The results for the default probability threshold cut (0.50) are shown in black, while we display in cyan the results for the optimal threshold cut (0.57).}
\label{fig:comp_pur_thre_opt_rad}
\end{figure}

Although we do not change the choice of the probability cut we show (Fig.~\ref{fig:comp_pur_thre_opt_rad}) the impact on C and P of this possible modification. Completeness (solid lines) and purity (dashed lines) are displayed as a function of radius, for two different probability cuts. The default cut (0.50) is in black, while the optimal cut (0.57) is in cyan. We can see that in any radius the gain in purity is not significant, while there is a large decrease in completeness for large radii ($> 0.75 \times R_{200}$). That reinforces our choice for not modifying the probability threshold. This figure is also complimentary to Figs. \ref{fig:comp_pur_magr}, \ref{fig:comp_pur_colour} and \ref{fig:comp_pur_clsmass}, as now we show the variation of C and P with radius. We see that C is $\ga$ 95\% for R$\la 0.5 \times R_{200}$, being $\sim$ 90\% in the the third bin and decreasing to $\sim$ 85\% in the outermost region. The variation in P is larger, from $\sim$ 94\% to $\sim$ 80\%, from the core to $\sim R_{200}$.

\subsection{Comparison to previous results} 
\label{comp_res}

A direct comparison to the previous efforts found in the literature \citep{geo11, bru00, roz15, cas16, bel18} is not straightforward. The main reason for that is the nature of the data employed in each case, the cluster mass and redshift ranges sampled, as well as the galaxy luminosity limits considered. For instance, \citet{geo11} covers a much larger redshift range ($0 < z < 1$) and is restricted to groups ($10^{13} < M_{200}/M_{\odot} < 10^{14}$), while we consider very low-$z$ ($z_{\text{phot-cl}} \le 0.045$) and with mean mass $\ga 5 \times 10^{14} M_{\odot}$. Nonetheless, it is worth summarizing their results. For a probability threshold of 0.5 they obtain C $=92$\% and P $=69$\% for their full sample. If we inspect their results according to different properties, we see that at low-$z$ (closer to us) they get similar results, C $\ga 90$\% and P $\sim 70$\%. However, as mentioned above, their sample comprises only groups (with masses $ < 10^{14} M_{\odot}$).

The work of \citet{cas16} is based on mock catalogs, with clusters in a very wide redshift range ($0.05 < z < 2.58$) and masses of $10^{13.29-14.80} M_{\odot}$. The mean purity and completeness they obtain are 56\% and 93\%, respectively. In their lowest redshift bin they obtain C $\sim 95$\% and P $\sim 70$\%. Note their results are based on a probability threshold cut of 0.2, thus favoring completeness at the cost of purity. They also consider only galaxies brighter than $H^* + 1.5$.

\citet{bru00} list the values of C and P for individual clusters, but not for the full stacked sample of galaxies. Their results are not shown as a function of redshift or cluster mass neither. In any case, their sample is composed of only nine clusters in a very different redshift range than ours ($z > 0.6$).

Some of the already mentioned works do not assess the performance through completeness and purity, making even harder a comparison. That is the case for the studies of \citet{bel18, roz15}. In the first case their main goal is cluster selection. The membership assignment is part of the process. In the second reference, they focus on selecting only red cluster galaxies. However, we can not make a direct comparison for this population as they do not give C and P values. Hence, we postpone a more detailed comparison with the literature to a future work, when we plan to consider groups and clusters, up to $z \sim 1$, from cosmological simulations.

\section{Applications}
\label{appl}

The reliable photometric membership selection of galaxies in clusters enables numerous astrophysical and cosmological applications. Without relying on a background correction one can investigate, for instance: the cluster luminosity function \citep{rib13b,pop06}, the spatial segregation of blue and red galaxies \citep{nas17}, the colour-magnitude relation \citep{gla98,cru04}, colour and morphology density relations \citep{dre80,lop14}, properties of transitional galaxies \citep{kan09,lop16}, substructure and magnitude gap of the first two brightest cluster galaxies \citep{lop06,tre17,lop18}, AGN fraction in clusters \citep{pim13,lop17}, scaling relations and mass calibration \citep{lop06,lop09b,pop05}. This technique can also be valuable for distinguishing cluster from background galaxies for lensing analysis \citep{mon17}.

\begin{figure}
\begin{center}
\leavevmode
\includegraphics[width=2.5in,angle=270]{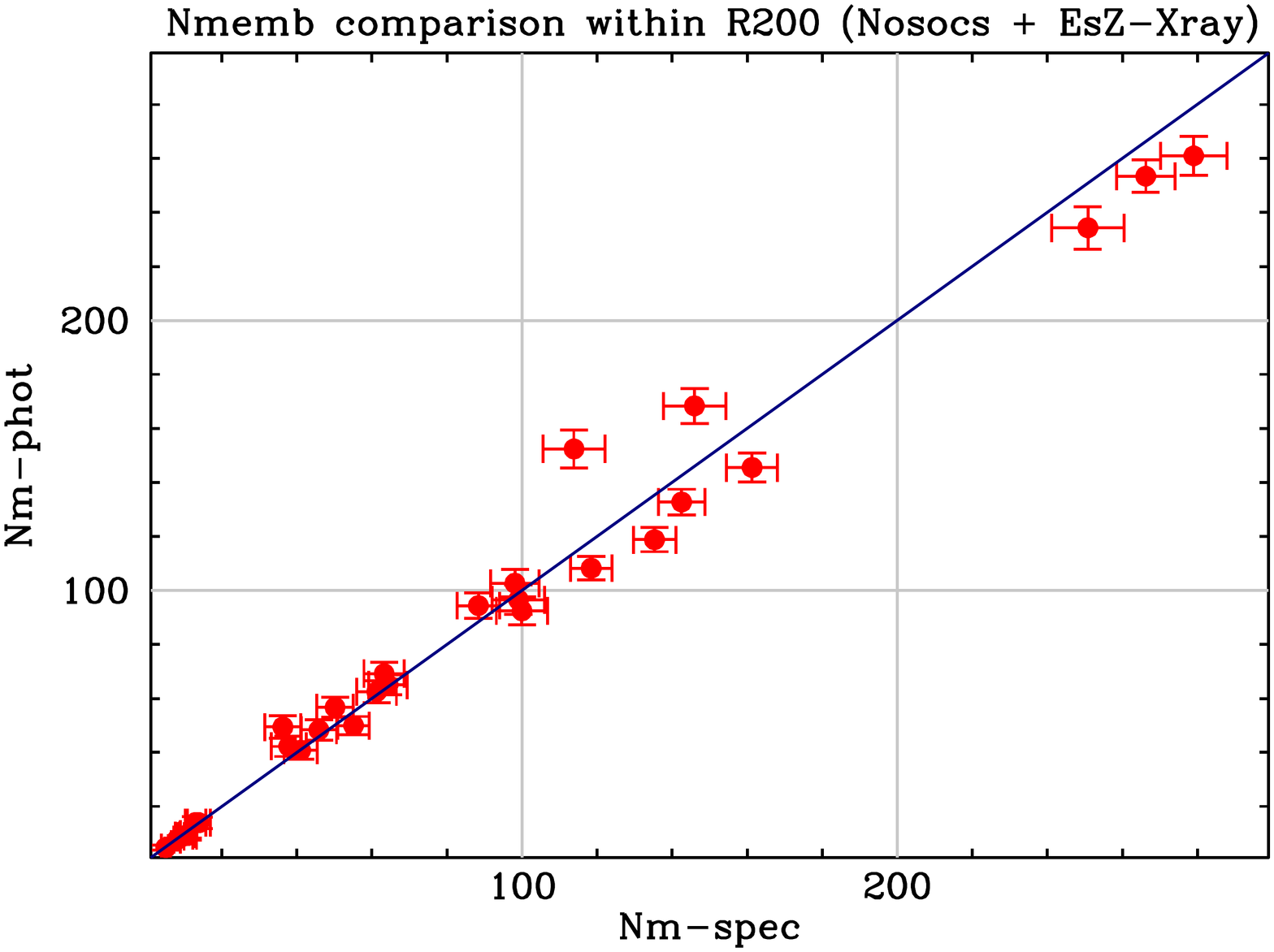}
\end{center}
\caption{The number of photometric galaxy members (within $R_{200}$ for $M_r \le M_r^* + 3.0$) {\it versus} the equivalent spectroscopic number.}
\label{fig:nmemb_spec_phot}
\end{figure}

We show below results of four different studies as examples of the power of our methodology. First, we show in Fig.~\ref{fig:nmemb_spec_phot} the basic test of the number of galaxies within the clusters. We show the number of photometric galaxy members {\it vs} the equivalent spectroscopic number, for $R \le R_{200}$ and $M_r \le M_r^* + 3.0$. The estimate in the Y$-$axis is derived by the sum of all the membership probabilities. It performs slightly better than simply counting the objects classified as members with a probability threshold set to 0.5. Note that we make the comparison with all galaxies in our sample. However, for mass calibration, perhaps we could test a different richness definition, considering brighter \citep{lop06} and maybe redder \citep{roz15} galaxies only. That could lead to a reduced scatter of scaling relations, which we will investigate in a future work.

In the meantime we compare the "classical" red sequence (RS) selection to what is achieved with our method. What we mean by "classical RS" is the simple selection of all galaxies in the red sequence as cluster members. Note that by no means we try to reproduce the selection of \citet{roz15}, who define a probability for a galaxy to be a RS cluster member. However, we try to mitigate the background contamination in the RS, making use of blank fields. Our procedure, for each cluster, is as follows. First we select all galaxies within $R_{200}$ and compute the mean magnitude (in the $r-$band, $r_{\text{mean}}$) of the first five brightest galaxies after the exclusion of the brightest. We also compute the mean (g-r) color of those five galaxies. Next we select the remaining galaxies having colors within $\pm 0.2$ of $(g-r)_{\text{mean}}$ and brighter than $r_{\text{mean}}+5.0$. We use those galaxies to obtain a linear regression of the RS. Next we select only galaxies within 2-$\sigma$ of the previous linear relation and perform a new fit. We repeat the last procedure until the linear regression solution converges (relative difference of both coefficients to their previous ones is less than 1\%). An example for the cluster Abell 2147 is show in Fig.~\ref{fig:abell_2147}.

\begin{figure}
\begin{center}
\leavevmode
\includegraphics[width=2.5in, angle=270]{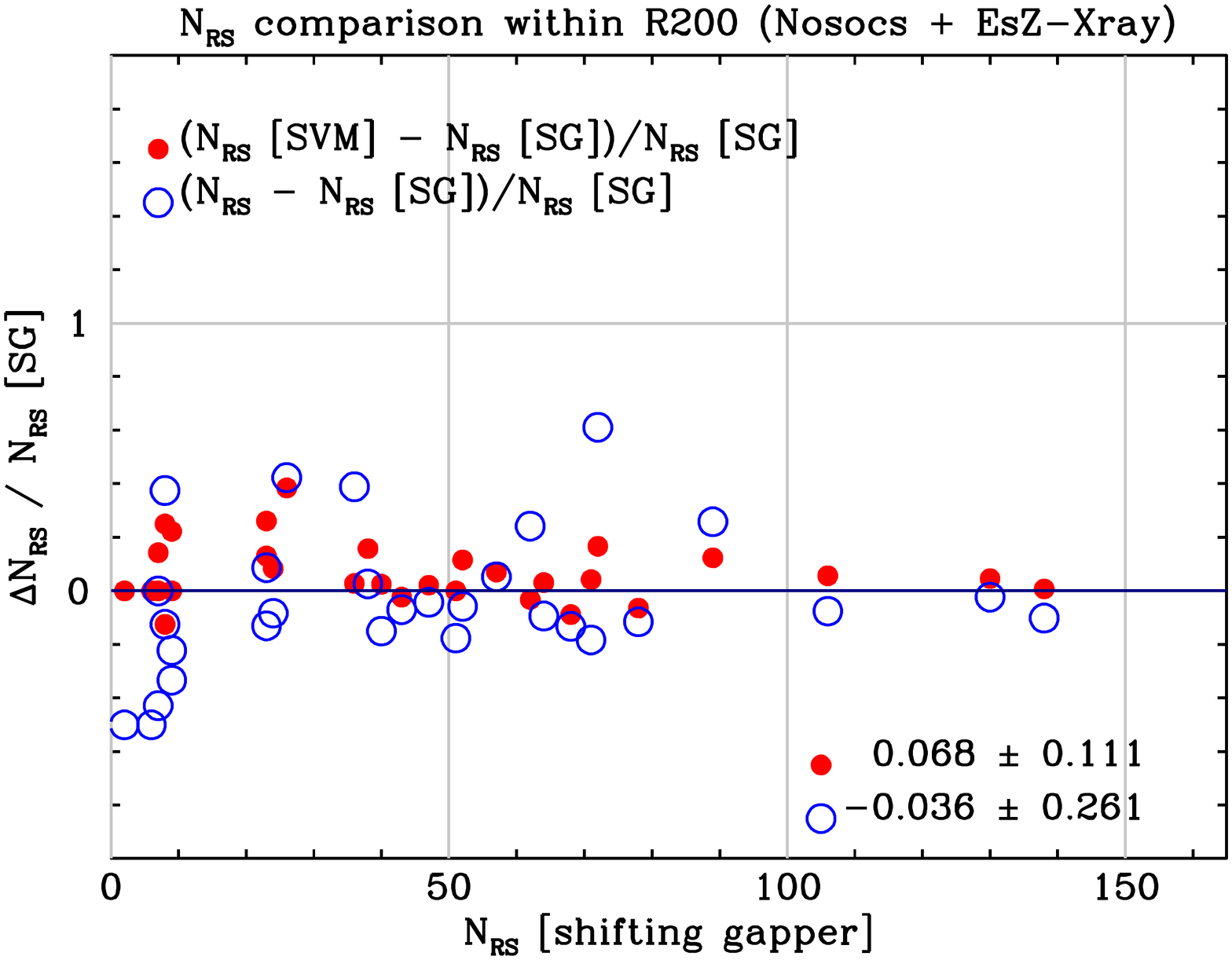}
\end{center}
\caption{The fractional difference between the number of red sequence galaxies (N$_{RS}$) obtained with a photometric approach and the equivalent number of spectroscopically selected galaxies, with the "shifting gapper" (SG) technique. Every point represents a cluster. The filled red points show the results considering the ML (SVM) method and the open blue symbols those obtained for the "classical RS" (simply summing up all galaxies within 2-$\sigma$ of the RS and applying a background correction). In the bottom right we show the mean fractional difference and standard deviation for each case.}
\label{fig:NRS_spec_phot}
\end{figure}

We compute the number of red sequence galaxies N$_{RS}$, simply counting the number of galaxies within 2-$\sigma$ of the final fit. We also impose a magnitude cut, considering only galaxies with $r_{\text{mean}} - 1.0 \le r \le r_{\text{mean}} + 3.0$. For the "classical RS" we take in account all galaxies (after background subtraction), while we consider only the ML members for our ML based estimate of N$_{RS}$. For the "classical RS" richness the background is estimated from 50 blank fields selected from the SDSS footprint \citep{lop07} and considers only galaxies within the same RS limits defined for each cluster. As a reference, we also count objects in the RS considering only galaxies that are spectroscopically selected members (from the "shifting gapper" technique). We show a comparison in Fig.~\ref{fig:NRS_spec_phot}, for our 30 clusters, of the photometric N$_{RS}$ values ("classical RS" and ML) to the spectroscopic values. The filled red points have the comparison for the ML counts, while the open blue symbols show the equivalent for the "classical RS". The mean fractional differences and standard deviation are listed in the bottom right. We verify there is only a small bias ($< 7\%$), both for the ML N$_{RS}$ values and the "classical RS". As expected, the scatter is larger for the "classical RS" approach. That is mainly due to the poorer systems.

\begin{figure}
\begin{center}
\leavevmode
\includegraphics[width=3.5in]{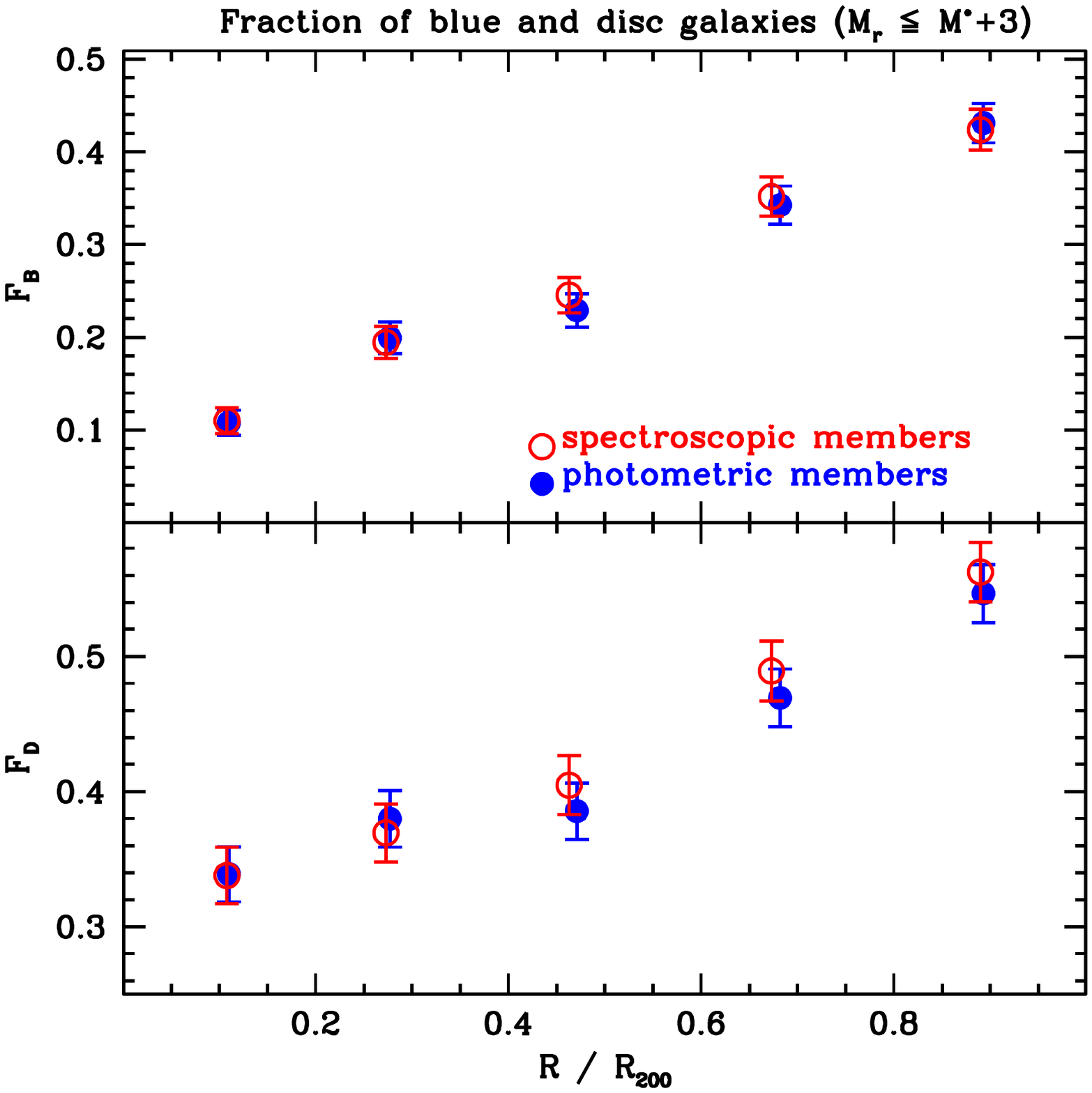}
\end{center}
\caption{Variation of the fractions of blue (top) and disc (bottom) cluster galaxies as a function of clustercentric distance. Results obtained from the spectroscopic member list are displayed in red open circles, while in blue we show the results for the photometric members.}
\label{fig:fb_fd_radbins}
\end{figure}

\begin{figure}
\begin{center}
\leavevmode
\includegraphics[width=2.5in, angle=270]{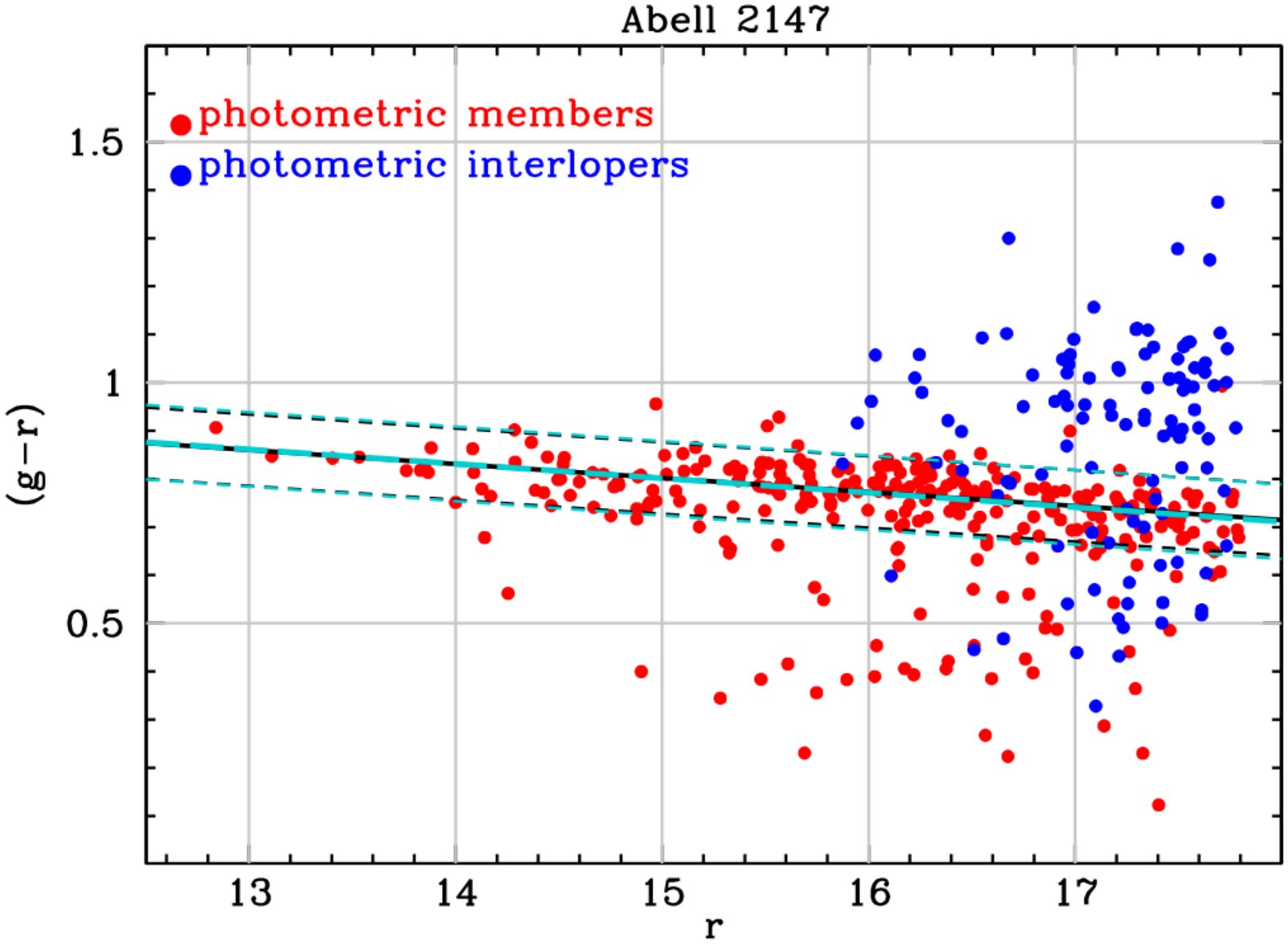}
\end{center}
\caption{Colour magnitude relation for the cluster Abell 2147. All galaxies within $R_{200}$ and $M_r \le M_r^* + 3.0$ are shown. Photometric selected members are in red, while the blue dots are for the galaxies photometric classified as interlopers. Two nearly coincident regression lines are displayed. In solid black the one obtained with red sequence galaxies from the photometric membership classification, in long dashed cyan using the spectroscopic members. In both cases, the 2-$\sigma$ limits are displayed with short dashed lines.}
\label{fig:abell_2147}
\end{figure}

We show in Fig.~\ref{fig:fb_fd_radbins} the fraction of blue (top panel) and disc (bottom) galaxies as a function of distance to the cluster center. As above, we use all member galaxies with $R \le R_{200}$ and $M_r \le M_r^* + 3.0$. However, now we take the list of members classified by assuming the membership probability threshold at 0.5. The blue and disc classification of galaxies is done according to \citet{lop14,lop16}. Blue galaxies are those with $(u-r) < 2.2$, while disc dominated have $C < 2.6$, where $C$ is the concentration index. The colour and morphological dependence with environment (indicated by the clustercentric distance) is evident. The agreement between the results derived from the photometric and spectroscopic members is remarkable. 

Finally, we show in Fig.~\ref{fig:abell_2147} the colour-magnitude relation [$(g-r)$ {\it vs} $r$] for Abell 2147. We exhibit the results for this cluster as it is the richest one in our sample. However, the main reason is the fact it has one of the largest discrepancies between the photometric and spectroscopic $R_{200}$ estimates ($\sim$ 35\%). $R_{200} = 1.84$ Mpc from the photometric scaling relation, while the original value is 2.84 Mpc from the virial analysis. Photometric selected members are displayed in red, while the interlopers are in blue. As before, we show all galaxies with $R \le R_{200}$ and $M_r \le M_r^* + 3.0$. An important point to highlight is the ability to recover not only red sequence (RS) galaxies, but also those in the blue cloud. Besides that we are also able to recognize and eliminate interlopers falling upon the RS. We also show two nearly coincident regression lines. In solid black the one obtained with red sequence galaxies from the photometric membership classification (ML), in long dashed cyan using the spectroscopic members. In both cases, the 2-$\sigma$ limits are displayed with short dashed lines. Both regression lines consider the galaxies displayed, within the photometric $R_{200}$ estimate. A third regression line (not shown) considering the spectroscopic members, but within the original $R_{200}$ value (from the virial analysis) is coincident with the other two. Even considering the larger original radius and the extra members between 1.84 and 2.84 Mpc, the colour-magnitude relation does not change.

Note the main goal of the current paper is not to derive a precise photometric $R_{200}$ estimate. We already have an approach with reasonable results from \citet{lop09b}, from which we get fractional differences always smaller than 40\% (for most clusters $<$ 20\%). After our final membership assignment we could actually use all the members to derive a new scaling relation between number of members and $R_{200}$, trying to improve the results from \citet{lop09b}, which had a richness estimate based on a statistical background correction. However, as mentioned above, to properly derive a scaling relation we should test the aperture used (e.g., 0.50 $h^{-1}$ Mpc), the luminosity range and perhaps color selection. As we have a reliable member list, with galaxy photometric redshift availables, we could also think on improving the cluster photometric redshift estimate (compared to the one from \citealt{lop07}). However, those applications are beyond the scope of the current work, which is to show a new approach to derive robust photometric membership estimates.

\section{Conclusions}

We developed a new method based on machine learning tecnhiques for estimating membership of galaxies in clusters, only from photometric information. All the features employed are derived solely from photometry, such as the $z_{\text{phot}}$, galaxy local densities, rest-frame magnitudes and colors, etc. 

The actual membership class is derived from spectroscopic data (employing the "shifting gapper" technique). The photometric parameters are then used as input for machine learning algorithms to predict their class (if member or not). 

Our data set is mostly composed of massive clusters (mean mass equal to $5.05 \times 10^{14} M_{\odot}$) in the local universe ($z_{\text{phot-cl}} \le 0.045$). We have 4420 galaxies (members and interlopers) within $R_{200}$ and with $M_r \le M_r^* + 3.0$, in 30 clusters. This sample is split in half, so that we can use 2210 galaxies for training purposes and the same number for validation of the results.

We investigate the best set of features for the photometric membership classification, choosing to work with a number of colors, magnitude, galaxy structural parameters and environmental tracers. Those are (u-r), (g-r), (g-i), (r-i), (r-z), r, LOG $\Sigma_5$, $R_{\text{petr}}$, $C$, $R/R_{200}$ and $z_{\text{phot}}$. After tuning the algorithms and selecting the best ones we find the SVM method to perform better, deriving completeness and purity values of C $\sim$ 92\% and P $\sim$ 87\%.

We further investigate possible dependencies in the performance related to galaxy magnitude, color and cluster parent mass. We find a small trend according to magnitude and a slightly more pronounced dependency with color, with bright and red galaxies displaying better results (higher C and P). On what regards the parent cluster mass, we detect a decrease in performance (traced by purity) with cluster mass, as P reaches lower values for lower masses. Then we used a group sample (containing less massive systems) confirming the trend with cluster mass, but detecting it also for completeness. However, we argue that other effects may impact those lower mass systems, such as miscentering and spectroscopic incompleteness. Hence, we postpone a more detailed analysis for a future work, based on mock catalogs.

We also show the variation of the results according to clustercentric distance, finding higher C and P in the central parts of clusters, when compared to larger radii. Finally, we discuss possible improvements if one chose to work with an optimal membership probability threshold (instead of the default value of 0.5). However, the results do not show a large variation.

In order to highlight the potential of our method we show a few applications. First, we perform the most basic test, verifying a very good agreement in the number of photometric selected members compared to the original spectroscopic values. That is very reassuring, as those clean photometric estimates of richness may provide a reliable way to estimate cluster mass for large samples of current and future sky surveys (such as DES, EUCLID and LSST). We also derive richness considering only RS galaxies, comparing two photometric counts (our ML ones and a "classical RS") to the spectroscopic RS richness. 

Next we show the color and morphological variation of galaxy populations within clusters, comparing the photometric and spectroscopic derived estimates. The excellent agreement stress how important this method can be for environment studies, based on cluster galaxies.

Finally, we show the color-magnitude relation for the richest cluster in our sample, obtained through a photometric selection of members and interlopers. The linear regression based on the photometric selected red sequence galaxies agrees extremely well with the one derived from the spectroscopic members. It is also reassuring to see the method is capable not only to select red cluster galaxies, but also objects in the blue cloud. Perhaps, even more important is the ability to exclude interlopers falling along the red sequence.

One can think of many other applications, such as the investigation of the cluster luminosity function, the relation of AGN and environment, lensing studies, etc (see an extended, but not complete list in $\S$\ref{appl}). We are currently working on the second paper of this series, on which we investigate the efficiency of our code with mock catalogs.

\section*{Acknowledgements}

PAAL thanks the support of CNPq, grant 309398/2018-5. ALBR thanks for the support of CNPq, grant 311932/2017-7. We acknowledge the anonymous referee for the very helpful suggestions. We are also thankful to A. Mahabal for the careful reading of the paper before its submission.

This research has  made use of the SAO/NASA  Astrophysics Data System, the NASA/IPAC Extragalactic  Database (NED) and the ESA Sky tool (sky.esa.int/).  Funding for the SDSS
and  SDSS-II was  provided  by  the Alfred  P.  Sloan Foundation,  the
Participating  Institutions,  the  National  Science  Foundation,  the
U.S.  Department  of  Energy,   the  National  Aeronautics  and  Space
Administration, the  Japanese Monbukagakusho, the  Max Planck Society,
and  the Higher  Education  Funding  Council for  England.  A list  of
participating  institutions can  be obtained  from the  SDSS  Web Site
http://www.sdss.org/. The 2dF Galaxy Redshift Survey (http://www.2dfgrs.net/) has been made possible by the dedicated efforts of the staff of the Anglo-Australian Observatory, both in creating the 2dF instrument and in supporting it on the telescope. We are also very thankful for the availability of the Final Release of 6dFGS (http://www-wfau.roe.ac.uk/6dFGS/).




\bibliographystyle{mnras}
\bibliography{rpmI.bib} 





\bsp	
\label{lastpage}
\end{document}